\documentclass[twocolumn]{aastex62}
\usepackage{apjfonts, graphicx, comment, xcolor}
\usepackage{multirow}

\begin{document}

\title{Modeling the upper kHz QPOs of 4U~1728$-$34 with X-ray reverberation}

\shorttitle{4U~1728$-$34}
\shortauthors{Coughenour et al.}

\author[0000-0003-0870-6465]{Benjamin~M.~Coughenour\affiliation{1}}
\author[0000-0002-8294-9281]{Edward~M.~Cackett\affiliation{1}}
\affiliation{Department of Physics \& Astronomy, Wayne State University, 666 W. Hancock St, Detroit, MI 48201, USA}

\author[0000-0002-0695-1662]{Philippe~Peille\affiliation{2}}
\affiliation{Centre National d'\'{E}tudes Spatiales, Centre Spatial de Toulouse, 18 Avenue Edouard Belin, 31401 Toulouse Cedex 9, France}

\author[0000-0002-9191-219X]{Jon~S.~Troyer}
\affiliation{Department of Physics \& Astronomy, Wayne State University, 666 W. Hancock St, Detroit, MI 48201, USA}

\email{coughenour@wayne.edu}

\begin{abstract}

While kilohertz quasi-periodic oscillations (kHz QPOs) have been well studied for decades since their initial discovery, the cause of these signals remains unknown, as no model has been able to accurately predict all of their spectral and timing properties. Separately, X-ray reverberation lags have been detected in AGN and stellar-mass black hole binaries, and reverberation may be expected to occur in neutron star systems as well, producing lags of the same amplitude as the lags measured of the kHz QPOs. Furthermore, the detection of a relativistically reflected Fe K line in the time-averaged spectra of many neutron star systems provides an additional motivation for testing reverberation. While it has been shown that the lag-energy properties of the lower kHz QPOs are unlikely to be produced by X-ray reverberation, the upper kHz QPOs have not yet been explored. We therefore model the upper kHz QPO lag-energy spectra using relativistic ray-tracing functions and apply them to archival \emph{RXTE} data on 4U~1728$-$34 where upper kHz QPOs have been detected. By modeling the time-averaged spectra in which upper kHz QPOs had been significantly detected, we determine the reflected flux fraction across all energies and produce a model for the lag-energy spectra from X-ray reverberation. We explore the dependence of the modeled lag properties on several different types of reflection models, but are unable to successfully reproduce the measured lags of 4U~1728$-$34. We conclude that reverberation alone does not explain the measured time lags detected in upper kHz QPOs.

\end{abstract}

\section{Introduction} \label{Intro}

Studying accretion around neutron stars (NSs) investigates both the extreme strong gravity environment as well as the physics of neutron stars themselves. The spectral and timing properties of NS low-mass X-ray binaries (LMXBs) have been studied intensively and provided several means of probing the inner accretion disk and the immediate environment of the neutron star. The detection of relativistic reflection features in the time-averaged spectra of NS LMXBs has made measurements of the innermost accretion disk possible \citep[e.g.][]{Bhattacharyya&Strohmayer2007, Cackett_etal2008, Cackett_etal2009a, Cackett_etal2010, Egron_etal2013, Miller_etal2013, diSalvo_etal2015, Sleator_etal2016, Ludlam_etal2017a, Ludlam_etal2019, Coughenour_etal2018}. The fastest timing signatures detected from NS LMXBs are the kilohertz quasi-periodic oscillations (kHz QPOs), and may also provide an independent glimpse into the physics in the vicinity of the neutron star and inner accretion region, though the physical mechanism(s) that produce kHz QPOs are not well understood \citep[see][for a review]{vanderKlis2000, vanderKlis2006}.

The launch of the \emph{Rossi X-ray Timing Explorer} (\emph{RXTE}) \citep{BradtRXTE1993} in 1995 made possible the discovery of high frequency variability in X-ray systems, and kHz QPOs were first detected shortly thereafter in NS LMXBs \citep[see, e.g.,][]{vanderKlis1998}. The frequencies of the kHz QPOs, roughly 300--1200 Hz, are of the order of the Keplerian orbital frequency of material at the inner disk boundary \citep{MillerMC_etal1998, Stella&Vietri1999}. This would suggest that the QPOs themselves are produced very near the neutron star, and are therefore a probe into the inner accretion environment. kHz QPOs are classified by their frequency as either the upper or lower kHz QPO, since in many systems twin kHz QPOs have been found simultaneously \citep[see, e.g.][]{vanderKlis_etal1997}. Aside from having different characteristic frequencies, the upper and lower kHz QPOs differ in a variety of ways. For example, the upper and lower kHz QPOs inhabit different regions on a plot of quality factor versus frequency, allowing the two to be distinguished in many cases when only one kHz QPO is detected \citep{Barret_etal2006}. The upper and lower kHz QPOs exhibit different spectral-timing properties as well \citep{deAvellar_etal2013, Peille_etal2015, Troyer_etal2018} suggesting a different emission mechanism between the upper and lower kHz QPOs.

Time lags between different energy bands have been measured utilizing kHz QPOs since they were first detected two decades ago \citep{Vaughan_etal1998, Kaaret_etal1999}. A series of more recent analyses, however, has measured the behavior of these lags across more energy bins, giving a lag-energy spectrum \citep{Barret2013, deAvellar_etal2013, deAvellar_etal2016, Peille_etal2015, Troyer&Cackett2017}. In fact, the entire \emph{RXTE} archive of NS LMXBs has been searched systematically for kHz QPOs, and lag-energy spectra measured for the lower kHz QPOs in 14 sources and for the upper kHz QPOs in just 6 sources, since these lags are more difficult to detect \citep{Troyer_etal2018}. This detailed survey of the upper and lower kHz QPOs has shown that their lag-energy spectra differ noticeably --- while the lower kHz QPOs typically show soft lags (the lower energy `soft' photons lag behind the higher energy `hard' photons) which steadily decrease at higher energies, the upper kHz QPOs tend to have a flat lag-energy spectrum that may increase at the highest energies, giving a hard lag \citep{Troyer_etal2018}. Any model which proposes to explain the emission of the kHz QPOs must also be able to explain the spectral-timing properties that have now been measured, including the lag-energy spectra.

These time lags could be caused by reverberation, though evidence for X-ray reverberation was first detected at an entirely different physical scale in active galactic nuclei (AGN) \citep{Fabian_etal2009, Zoghbi_etal2010, Zoghbi_etal2011}. More recently, an Fe K time lag was measured for NGC 4151 by \citet{Zoghbi_etal2012} \citep[see][for a review]{Uttley_etal2014}. X-ray reverberation lags have since been shown to be a powerful tool for studying AGN \citep{DeMarco_etal2013a, Alston_etal2014, Alston_etal2015, Cackett_etal2014, Kara_etal2014, Kara_etal2015, Kara_etal2016, Zoghbi_etal2014, Wilkins_etal2017,  Mallick_etal2018}. Thermal reverberation has also been recently detected in black hole X-ray binaries \citep{Uttley_etal2011, DeMarco_etal2015}, and in the case of the transient black hole source MAXI J1820+070 an Fe line lag is detected as well \citep{Kara_etal2019}. For AGN, it has been shown that X-ray reverberation time lags scale roughly with mass \citep{DeMarco_etal2013a, Kara_etal2013}, and that scaling relation provides a reasonable estimate for the lags detected for the black hole binaries (at much smaller mass scales) as well, since some measured lags in binary systems are longer than would expected by scaling with mass \citep{DeMarco_etal2013b}, and yet others are shorter \citep{Kara_etal2019}. Scaling the reverberation time lags measured in AGN to 1.4 $M_\odot$ gives a time lag on the order of 10 $\mu$s, which is the magnitude of the time lags measured at QPO frequencies for LMXBs \citep{Cackett2016}.

Since the reflection spectrum is a regularly occurring feature in neutron star LMXB spectra, one would expect reverberation to naturally occur in these systems due to the light travel time between the incident X-ray source and the irradiated disk \citep{Stella1990, Campana&Stella1995}. Considering that the frequencies of the kHz QPOs correspond to activity very near the neutron star --- particularly the upper kHz QPO which has a frequency which may correspond to the orbit of the inner accretion disk --- as well as the expected reverberation time lags in comparison with the measured lags, testing whether reverberation might explain the kHz QPO lags is essential. Such a test was recently carried out by \citet{Cackett2016} for the lower kHz QPO lags of 4U~1608$-$52, demonstrating that reverberation alone cannot reproduce the measured lag-energy spectrum. However, reverberation has never been tested against the upper kHz QPOs, which is the motivation of this work. In particular, \citet{Cackett2016} predicted hard lags increasing at higher energies in the lag-energy spectrum, and since hard lags have been detected for several of the upper kHz QPOs \citep[e.g.,][]{deAvellar_etal2013, Peille_etal2015}, the upper kHz QPOs lags appear more promising for reverberation.

Among the NS LMXBs with detectable upper kHz QPOs, 4U~1728$-$34 has more upper kHz QPO detections and the best photon statistics to measure spectral-timing properties, and this source was investigated in depth by \citet{Peille_etal2015}. Furthermore, a strong reflection component has recently been detected in 4U~1728$-$34 by \citet{Sleator_etal2016, Mondal_etal2017, Wang_etal2019}, and a host of earlier observations show a broad Fe line to be present \citep{diSalvo_etal2000, Piraino_etal2000, Ng_etal2010, Egron_etal2011, Seifina&Titarchuk2011, Tarana_etal2011}. It is therefore the ideal source to investigate X-ray reverberation modeling.

We model the lag-energy spectrum using reverberation for a neutron star and compare it with the lag-energy spectra of the upper kHz QPOs in 4U~1728$-$34 as measured by \citet{Peille_etal2015}. As our aim is to model the lag-energy spectrum, an important part of that process is to determine the strength of the reflection spectrum across 3--25 keV. We describe our data selection and spectral analysis of 4U~1728$-$34 in Sections~\ref{Data} and \ref{Spectra}, and then we discuss our reverberation modeling and results in Section~\ref{Model}. We compare our results with the measured lags for 4U~1728$-$34, and discuss the implications in Section~\ref{Discuss}. Finally, we conclude our findings and their significance in Section~\ref{Conclusions}.

\begin{deluxetable}{lll}[t]
\tablecolumns{3}
\tablecaption{Exposure Times for Each QPO Frequency Group}
\tablehead{QPO Frequency Group & Number of Spectra & Total Exposure Time}
\startdata
700--800 Hz & 15 & 43813 \\
800--900 Hz & 14 & 50948 \\
900--1000 Hz & 6 & 20350 \\
1000--1100 Hz & 1 & 2009 \\
\enddata
\tablecomments{Each individual spectrum has a different exposure time, which is no less than $\sim$2000~s. Total exposure times are given in seconds, and provide the sum of exposures for each of the spectra within a frequency group. Detections of an upper kHz QPO above 1000~Hz were rare, and it should be noted that this frequency group has comparably worse statistics than the others.}
\end{deluxetable}

\section{Data reduction \& analysis} \label{Data}

We take the lag-energy data calculated by \citet{Peille_etal2015} for the upper kHz QPOs detected for 4U~1728$-$34. \citet{Peille_etal2015} used all archival \emph{RXTE} observations of 4U~1728$-$34 and separated each observation into 1024~s power spectra in order to detect kHz QPO signals. Upper kHz QPOs are unambiguous when detected simultaneously with a lower kHz QPO, while the QPO quality factor and frequency together can be used to distinguish between an upper and lower kHz QPO for individual detections \citep{Barret_etal2006}. The lag-energy spectrum describes the lag calculated between each energy bin and the reference band (3--25 keV) with that specific bin's light curve subtracted. See \citet{Nowak_etal1999} and \citet{Uttley_etal2014} for detailed discussions on calculating lags. Relatively large energy bins were used, with 1024~s data segments grouped by QPO peak frequency into four separate groups or bins: 700--800 Hz, 800--900 Hz, 900--1000 Hz, and 1000--1100 Hz for the upper kHz QPO. By separating the upper kHz QPO detections into different frequency groups, an attempt can be made to determine whether the lag-energy spectrum may depend on QPO frequency, since the peak frequency of the kHz QPOs changes over time and between observations. For more details regarding the data extraction please see \citet{Peille_etal2015}.

We therefore take a time-averaged spectrum for each ObsID where at least two upper kHz QPOs were detected (that is, where two or more 1024~s segments contained an upper kHz QPO detection). This was done to maximize the signal-to-noise of each spectrum while also allowing for the possibility that spectral differences occur between different ObsIDs or at different QPO frequencies. In doing so, we have a list of 36 \emph{RXTE} PCA time-averaged spectra, each containing roughly 2048 seconds or some greater multiple of 1024 seconds of exposure time (slight differences in exposure times between individual spectra are a result of the removal of X-ray bursts and count rate drops from the data). These spectra are then grouped depending on the frequency of the upper kHz QPO detected. This results in 15 individual spectra for the 700--800 Hz bin, 14 spectra in the 800--900 Hz bin, 6 spectra in the 900--1000 Hz bin, while just one spectrum which falls in the 1000--1100 Hz bin. A summary of this information is given in Table~1, along with the total exposure time for the available spectra in each QPO frequency group.

\begin{figure}[htb]
\includegraphics[trim=0.5cm 1.3cm 0cm 2.5cm, clip=true, width=10cm]{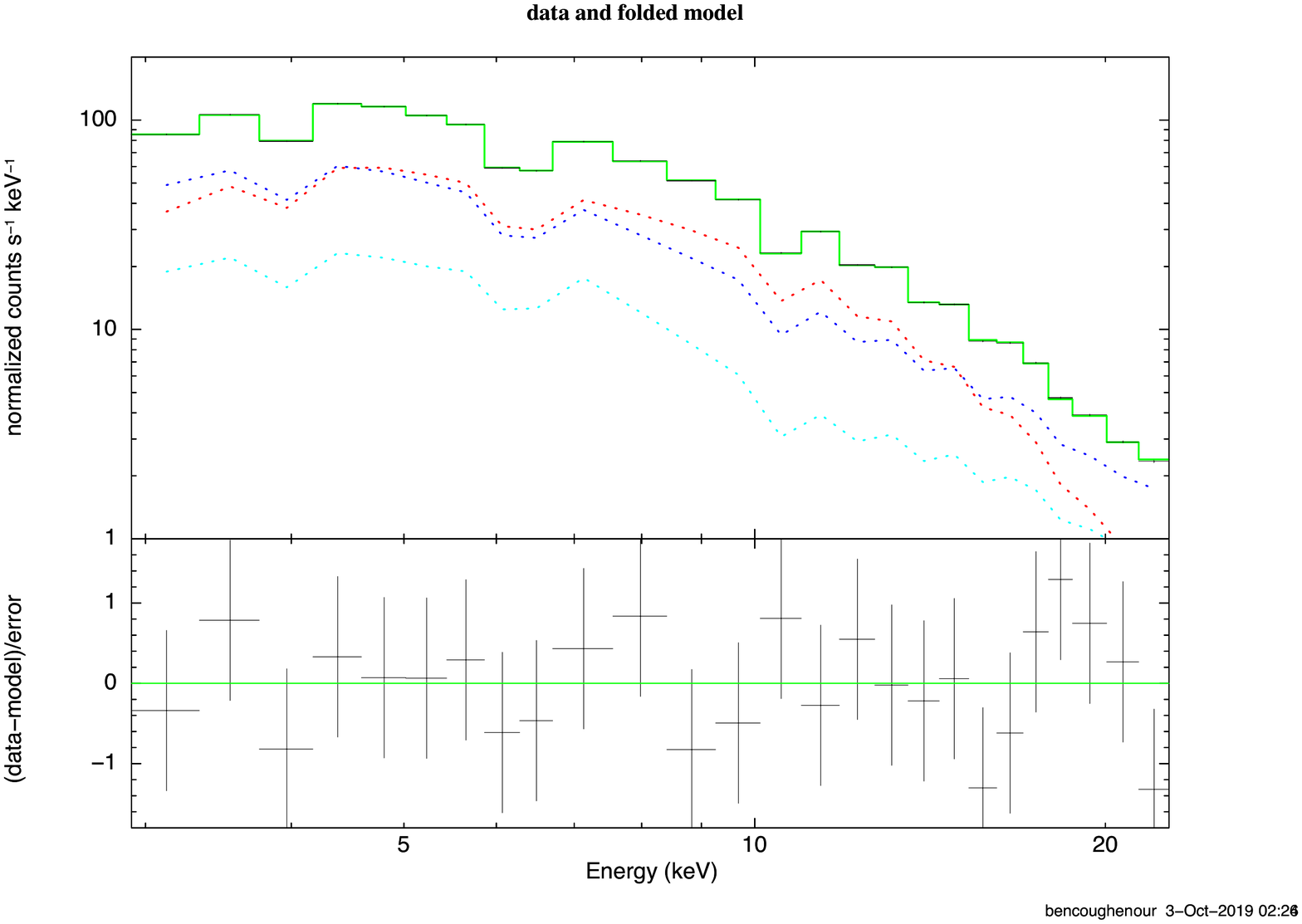}
\includegraphics[trim=0.5cm 1.3cm 0cm 2.5cm, clip=true, width=10cm]{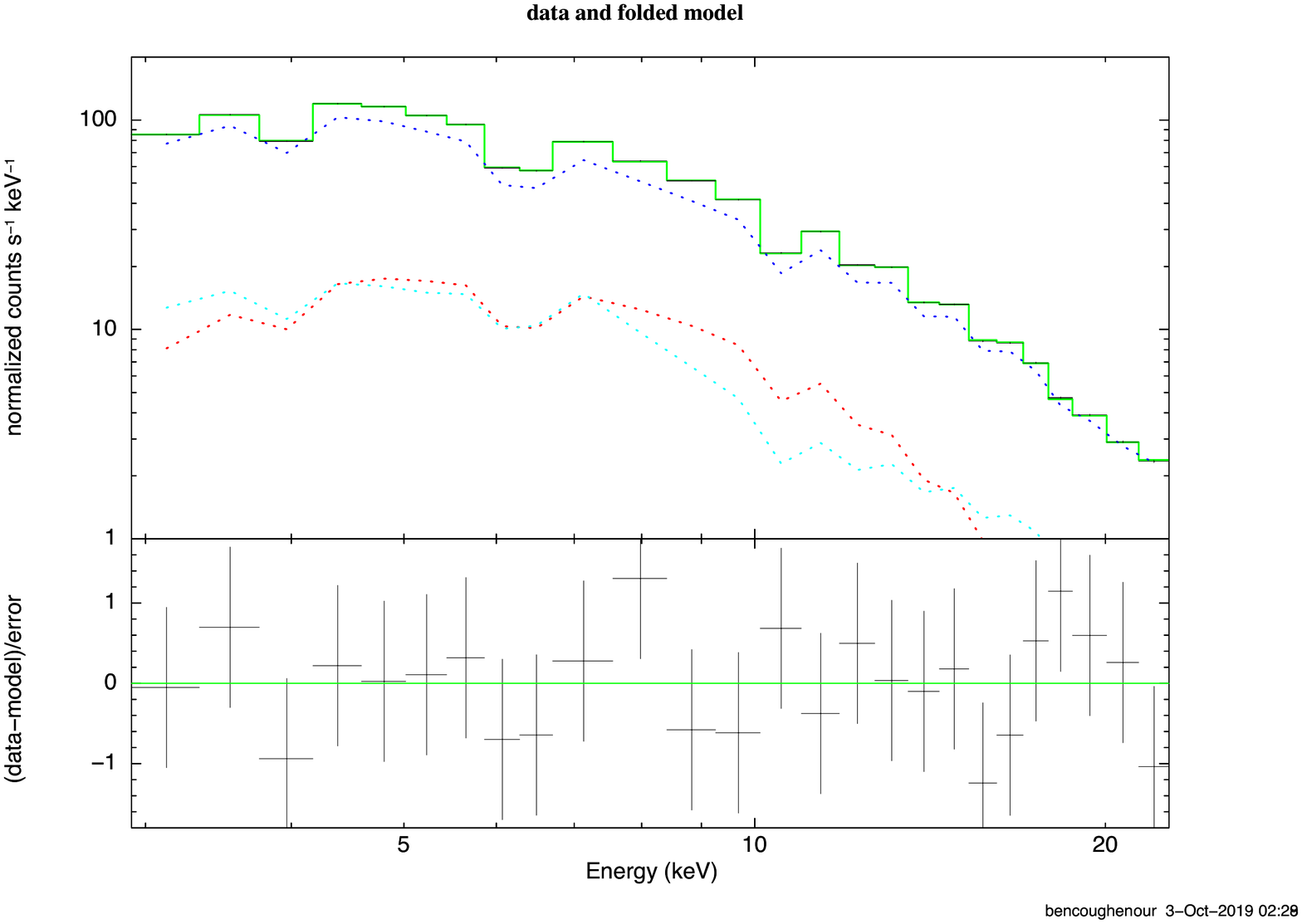}
\caption{Example of an individual spectrum (ObsID: 40019-03-01-00, 3071~s exposure) and two of our best-fit models using \texttt{relxill}, with normalized residuals given below each spectrum. We include either a multicolor disk blackbody (upper) or single temperature blackbody (lower) component along with the incident power law and reflection spectrum. In both figures, green represents the model flux, while the red dashed line is the thermal component (either \texttt{diskbb} or \texttt{bbodyrad}), and the blue and cyan dashed lines represent the incident power law and reflected flux, respectively. Both models give a reasonable fit to the limited data, with a reduced $\chi^2 \sim$ 0.6.}
\label{spectrum}
\end{figure}

\section{Spectral Modeling} \label{Spectra}

To model the lag-energy spectrum for the upper kHz QPOs of 4U~1728$-$34, the entire reflection spectrum must be taken into account. We therefore model the 36 individual time-averaged spectra, each representing its own ObsID where the upper kHz QPO was detected, using XSPEC v. 12.9.1 \citep{Xspec}. Each spectrum is limited to the 3--25 keV energy range, where the signal-to-noise ratio of the QPO detection is strongest and the spectra are still well above the background (which dominates above 25 keV). A systematic error of 0.6$\%$ was added to each channel for the PCA spectra.

We include neutral Hydrogen absorption using the model \texttt{tbabs} with photoionization cross sections and abundances set to the values found in \citet{Wilms_etal2000}. The neutral Hydrogen column density was fixed to a value of $4.0\times10^{22}$~cm$^{-2}$, based on the most recent spectral analyses of 4U~1728$-$34. Recent studies using simultaneous observations with \emph{Swift} and \emph{NuSTAR} have measured $N_{\rm H}$ to be between 3.9 and 4.5 $\times10^{22}$ cm$^{-2}$ \citep{Sleator_etal2016, Mondal_etal2017}, and a value of $4.5\times10^{22}$ was also reported by \citet{Wang_etal2019} using \emph{XMM-Newton} observations. While earlier studies of 4U~1728$-$34 reported considerably lower values \citep[see, e.g.][]{DAi_etal2006, Egron_etal2011}, we find that above 3 keV these differences are small, and that the \emph{RXTE} spectra do not satisfactorily constrain the absorption column density.

In modeling X-ray reverberation we assume that reflection occurs, and are also motivated by the fact that reflection features have been detected in 4U~1728$-$34 before. Measurements of the reflection spectrum of 4U~1728$-$34 have been done recently using simultaneous \emph{NuSTAR} and \emph{Swift} observations \citep{Sleator_etal2016, Mondal_etal2017}, and with  \emph{XMM-Newton} as well \citep{Wang_etal2019}. Earlier observations of 4U~1728$-$34 with a number of different X-ray telescopes showed an Fe line to be clearly present \citep{diSalvo_etal2000, Piraino_etal2000, DAi_etal2006, Ng_etal2010, Egron_etal2011, Seifina&Titarchuk2011, Tarana_etal2011}. We therefore aim to fit the \emph{RXTE} spectra for 4U~1728$-$34 with physical reflection models that will take into account not just the Fe line, but the full reflection spectrum, including the Compton hump above 10 keV. Such an approach is needed in order to model reverberation.

\begin{deluxetable*}{lLLLLLLLLL}[hbt!]
\centering
\tabletypesize{\footnotesize}
\tablecolumns{10}
\tablewidth{0pt}
\tablecaption{Best Fit Spectral Model Results by QPO Group}
\tablehead{QPO Group & \multicolumn{9}{l}{Model Component Parameters}}
\startdata
& \multicolumn{2}{l}{\texttt{diskbb}} & \multicolumn{7}{l}{\texttt{relxill}} \\[2pt]
			& \rm{T$_{in}$} \: \rm{ (keV)} & \rm norm 	& \rm{R$_{in}$ \: (ISCOs)} & $\Gamma$ &  \rm{log} \: $\xi$ & \rm{E$_{cut}$ \: (keV)} & \rm refl. \: frac. & \rm norm \: ($10^{-3}$) & $\chi^2$ \: \rm{(dof)} \\[2pt]
700--800 Hz 	& 3.40 \pm0.03 & 0.70^{+0.03}_{-0.02} 	& 2.8^{+1.2}_{-0.5} & 2.18^{+0.01}_{-0.05} & 3.23^{+0.05}_{-0.04} & 30.6^{+36.2}_{-6.6} & 0.27 \pm0.02 & 6.1^{+0.1}_{-0.2} & 9.51 (19.47) \\[2pt]
800--900 Hz 	& 3.23 \pm0.02 & 0.89 \pm0.03 		& 2.9^{+1.6}_{-0.9} & 2.37^{+0.01}_{-0.04} & 3.32^{+0.07}_{-0.04} & 567^{+244}_{-14} & 0.28^{+0.02}_{-0.03} & 6.2 \pm0.2 & 12.48 (19.14) \\[2pt]
900--1000 Hz 	& 3.16 \pm0.03 & 1.09^{+0.05}_{-0.04} 	& 15.4^{+29.1}_{-2.9} & 2.55^{+0.03}_{-0.06} & 3.52^{+0.15}_{-0.10} & 65.5^{+144.6}_{-13.5} & 0.28^{+0.05}_{-0.03} & 6.8 \pm0.5 & 9.84 (18.83) \\[2pt]
1000--1100 Hz 	& 3.12^{+0.04}_{-0.05} & 1.71 \pm0.10 	& 32.2^{+\rm{limit}}_{-27.8} & 3.17^{+0.05}_{-0.23} & 4.21^{+0.45}_{-0.42} & 1000^{+\rm{limit}}_{-964} & 0.70^{+0.57}_{-0.27} & 27.9^{+7.3}_{-7.6} & 9.77 (21) \\[4pt]
\hline
& \multicolumn{2}{l}{\texttt{bbodyrad}} & \multicolumn{7}{l}{\texttt{relxill}} \\[2pt]
			& \rm{kT \: (keV)}	& \rm norm 				& \rm{R$_{in}$ \: (ISCOs)} & $\Gamma$ &  \rm{log} \: $\xi$ & \rm{E$_{cut}$ \: (keV)} & \rm refl. \: frac. & \rm norm \: ($10^{-3}$) & $\chi^2$ \: \rm{(dof)} \\[2pt]
700--800 Hz 	& 2.12^{+0.03}_{-0.04} & 2.44^{+0.20}_{-0.16} 	& 1.6^{+0.6}_{-0.3} & 1.68^{+0.05}_{-0.03} & 3.19^{+0.10}_{-0.07} & 14.4^{+2.2}_{-0.9} & 0.14 \pm0.01 & 6.4 \pm0.2 & 9.16 (19.47) \\[2pt]
800--900 Hz 	& 2.19 \pm0.02 & 3.04^{+0.19}_{-0.13} 		& 3.1^{+2.9}_{-0.4} & 1.88^{+0.04}_{-0.05} & 3.19^{+0.12}_{-0.07} & 15.0^{+2.3}_{-1.0} & 0.14 \pm0.01 & 6.9^{+0.3}_{-0.2} & 11.39 (19.14) \\[2pt]
900--1000 Hz 	& 2.20 \pm0.03 & 3.26^{+0.28}_{-0.24} 		& 4.7^{+10.0}_{-1.4} & 1.83 \pm0.07 & 3.30^{+0.16}_{-0.14} & 10.3^{+1.5}_{-0.9} & 0.15^{+0.02}_{-0.01} & 7.2 \pm0.5 & 9.15 (18.83) \\[2pt]
1000--1100 Hz 	& 2.25^{+0.06}_{-0.05} & 4.80^{+0.65}_{-0.61} 	& 8.4^{+\rm{limit}}_{-5.1} & 1.98^{+0.16}_{-0.14} & 3.29^{+0.48}_{-0.26} & 10.0^{+2.8}_{-1.4} & 0.22 \pm0.06 & 9.5^{+1.0}_{-0.8} & 8.09 (21) \\[2pt]
\tableline
\enddata
\tablecomments{Results shown are the weighted average of fitting all of the spectra from each QPO frequency group with the models shown. Temperatures for \texttt{bbodyrad} and \texttt{diskbb} are given in keV, as is the high energy cutoff for \texttt{relxill}. Errors are given as 90 \% confidence intervals. Where `limit' is given as the value for the error, that parameter is not constrained to be less than the upper limit of the \texttt{relxill} model, which is 100 ISCOs in the case of $R_{in}$ and 1000 keV in the case of E$_{cut}$. This can be seen in the case of the individual spectrum which represents the 1000--1100 Hz frequency group.}
\end{deluxetable*}


The continuum of NS LMXB spectra are typically fit with a combination of several different model components. A power law is often used to represent the Comptonized continuum, which is the Compton upscattering of low-energy thermal photons from the accretion disk --- or thermal photons from the neutron star or boundary layer \citep{Gilfanov&Revnivtsev2005} by a cloud of hot electrons. One or more thermal components are then typically added. That may be a multi-temperature blackbody peaking in the soft X-rays to represent the accretion disk, or a single temperature blackbody to represent either the surface of the neutron star or the boundary layer between the neutron star and the accretion disk \citep{Popham&Sunyaev2001}. For a discussion on continuum modeling, see \citet{Lin_etal2007}.

We first fit the individual spectra with the simplest model, neutral absorption of a power law with a high-energy cutoff, using the model \texttt{tbabs}*\texttt{cutoffpl}. This provided an extremely poor fit to the data, with $\chi^2/$dof$~=~393.3/24.3$ when averaged over all 36 spectra. With the addition of either a thermal multicolor disk or blackbody (\texttt{diskbb} and \texttt{bbodyrad} in \textsc{xspec}, respectively), we found a dramatic improvement of fit, yielding either $\chi^2/$dof$~=~200.2/22.3$ or $173.2/22.3$, respectively. The residuals from these fits were dominated by a broad Fe line at $\sim$ 6 or 7 keV, and so the addition of a Gaussian Fe line was tested. This dramatically improved the fits, providing $\chi^2/$dof$~=~28.8/19.3$ for the disk model and $\chi^2/$dof$~=~27.0/19.3$ for the blackbody model. While an improvement, these models do not provide a satisfactory fit. Furthermore, considering the goal of calculating a lag-energy spectrum due to reverberation, we next try a more physically consistent reflection model.

When reflection is present, the reflection spectrum must be in response to either the Comptonized emission or the higher energy thermal emission from the neutron star or boundary layer. We begin with the \texttt{relxill} model for reflection from an incident power law with a high energy cutoff, \citep{DauserGarcia_etal2014, GarciaDauser_etal2014}, and include either a blackbody or multicolor disk thermal component. These model components, \texttt{diskbb} and \texttt{bbodyrad}, both include as variable parameters a temperature and normalization. The parameters of \texttt{relxill} are the emissivity index of disk, the dimensionless spin parameter $a$ for the neutron star or black hole, the inclination of the disk, the inner and outer disk radii $R_{in}$ and $R_{out}$, a redshift value $z$ for distant sources, the index of the incident power law spectrum $\Gamma$, the logarithm of the ionization parameter $\xi$, the disk's Fe abundance, a high-energy cutoff for the power law, the reflection fraction, and the normalization.

For the emissivity index, \texttt{relxill} allows a broken power law for the emissivity with changes in disk radius, with two power law indices and a `breaking' radius between them. We fix the index with classical value of 3 throughout the disk to simplify our model. This is reasonable considering the limited spectral resolution of \emph{RXTE} PCA. The neutron star spin rate of 4U~1728$-$34 has been measured previously via burst oscillations to be 363 Hz \citep{Strohmayer_etal2001}, which gives a dimensionless spin parameter $a = cJ/GM^2 = 0.15$ and so we use that value in our analysis. Since important parameters for the relativistic blurring are set by the shape of the reflection features (e.g., the Fe line), we use parameters determined using higher resolution spectra from \emph{NuSTAR}. Thus, we fix the disk inclination to 32 degrees, the value found using \emph{NuSTAR} and \emph{Swift} observations by \citet{Mondal_etal2017}, which is consistent with the values found in \citet{Sleator_etal2016}. We fixed the outer disk radius to 1000 $R_G$ (gravitational radii, $R_G = GM/c^2$) and we fixed the redshift to zero, as 4U~1728$-$34 is a Galactic source. Finally, we fix the Fe abundance equal to the solar abundance, since it is not well constrained by our spectra. All other parameters were allowed to vary.

\begin{figure*}[hbt]
\includegraphics[trim=3.0cm 4.2cm 2.5cm 5.2cm, clip=true, width=9cm]{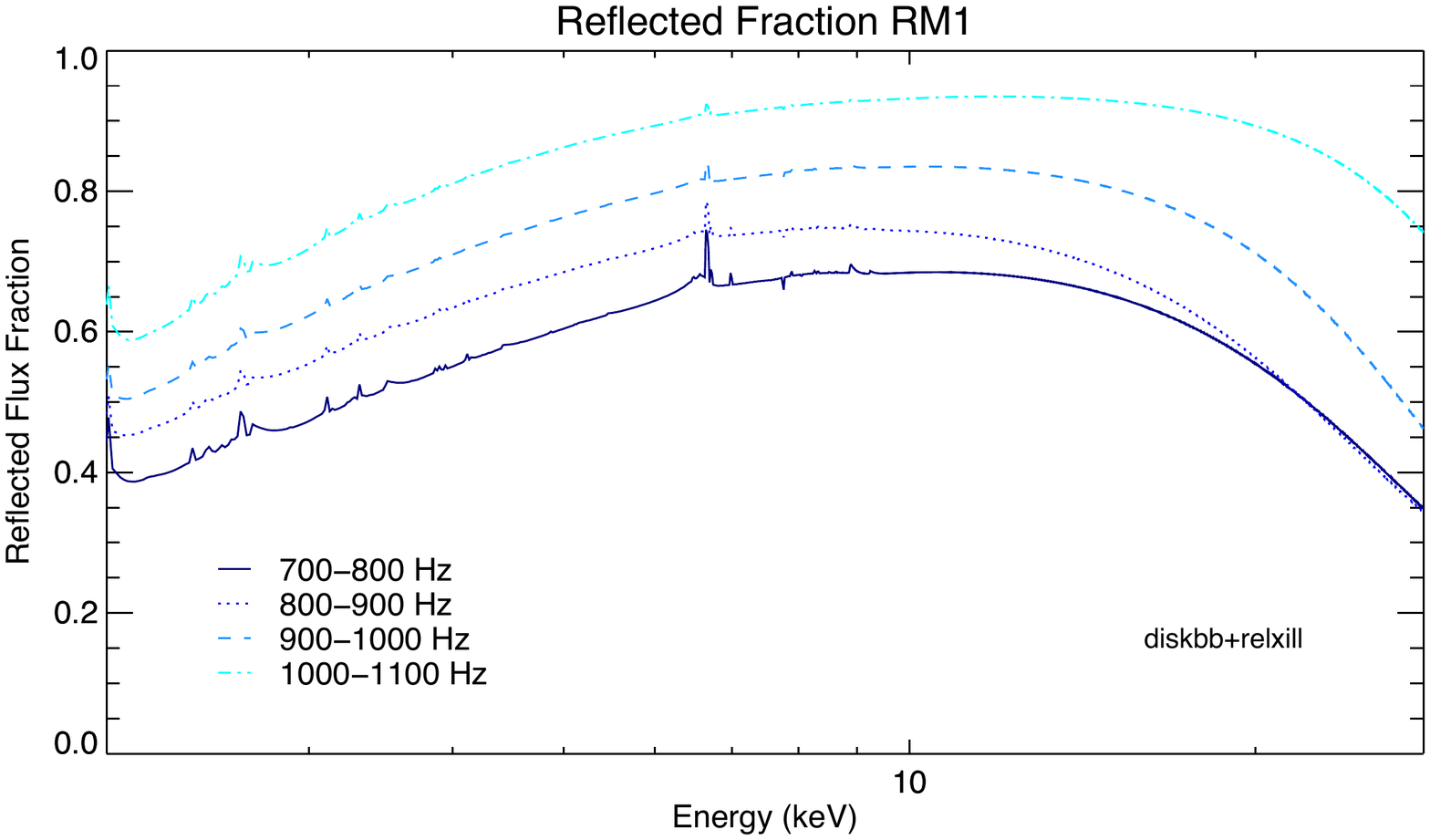}
\includegraphics[trim=3.0cm 4.2cm 2.5cm 5.2cm, clip=true, width=9cm]{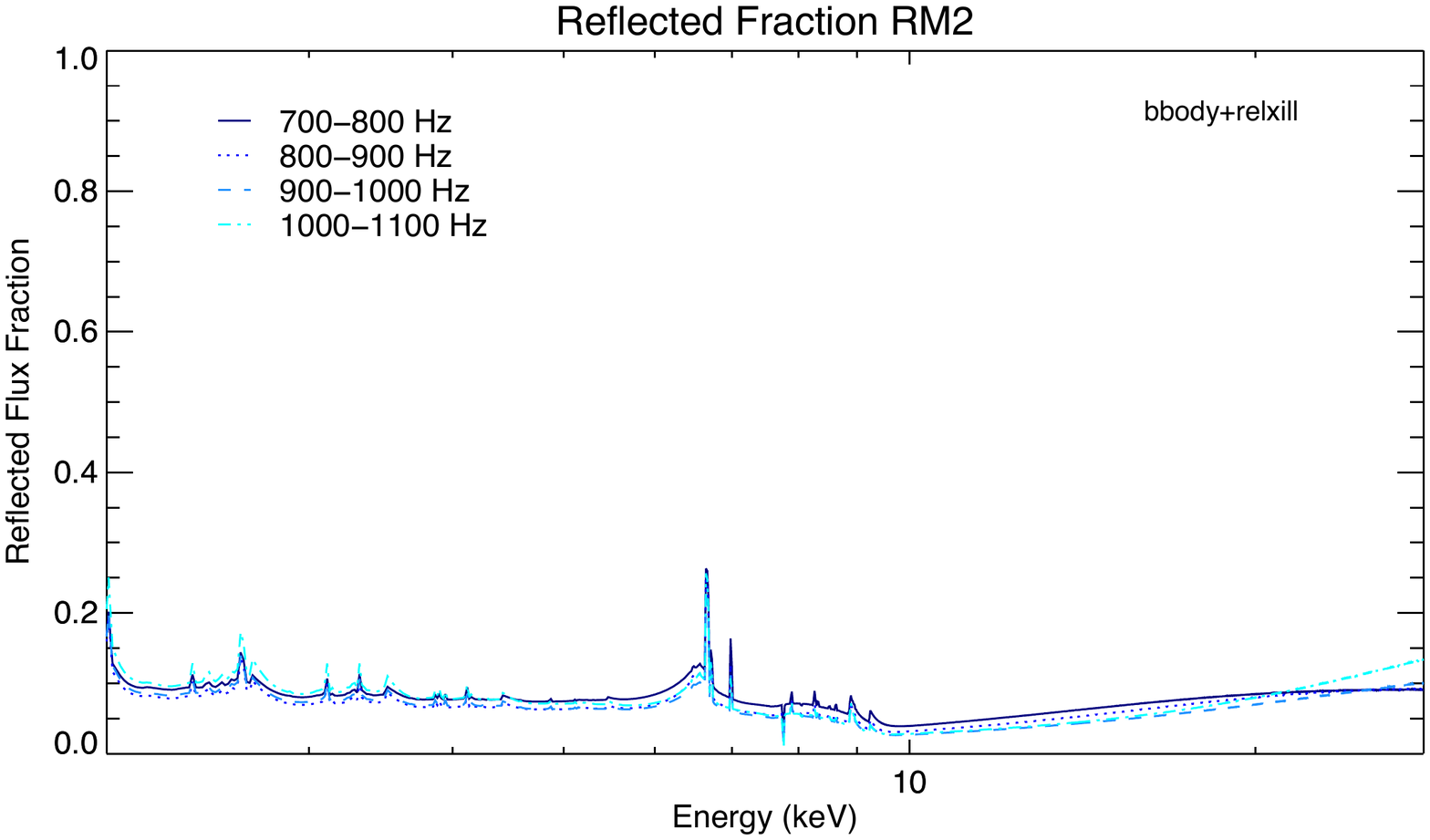}
\caption{Reflected fraction plotted against energy for both the disk (left) and blackbody (right) models, for each QPO frequency group. The reflected fraction is the ratio of the reflection spectrum to the total flux, and should not be confused with the reflected fraction parameter from the \texttt{relxill} model. In the case of the disk models, flux from the disk is included as thermal reverberation, and the disk shape and Compton hump provide most of the reflected emission. For the case of the blackbody, the reflection spectrum is somewhat diluted at lower energies and shows strong Fe line features and a Compton hump above 10 keV.}
\label{refl_frac}
\end{figure*}

We initially attempted \texttt{relxill} along with the addition of both a disk and a blackbody, however for most of the 36 spectra our best-fit results had a near-zero normalization for one of those two thermal components, and the quality of the fits were not improved against models using either a disk or a blackbody alone. We therefore analyzed \texttt{relxill} with the inclusion of either a disk or blackbody rather than the combination of the two. Since these two thermal continuum models are degenerate with the limited statistics of our 36 individual spectra, we continue to use both models in order to test the dependence of the lags on the choice of continuum. While there are only slight differences between the spectral results when considering both models, the fact that a disk might exhibit thermal reverberation will be important when modeling the lags. We refer to the \texttt{relxill+diskbb} model and to the \texttt{relxill+bbodyrad} model as the `disk' and `blackbody' models, respectively, throughout the remainder of the paper. Figure~\ref{spectrum} shows one spectrum fit by both models as an example, with normalized residuals. These models produce fit statistics of $\chi^2/d.o.f. < 1.0$ for almost every spectrum, with an average fit statistic of $\chi^2/d.o.f. \approx 0.5$ for both models. Although this is not ideal and values of $\chi^2$/d.o.f. closer to 1 might be obtained for these spectra using a simpler model, our goal is to test reverberation as a mechanism to produce the measured kHz QPO lags. We therefore continue our analysis using \texttt{relxill}. In doing so we are already slightly overfitting the \emph{RXTE} time-averaged spectra, and yet the addition of a thermal component is still required to achieve an adequate fit because \texttt{relxill} is unable to do so alone. We also tested the \texttt{relxillCp} model, which uses as its incident spectrum a more physical Comptonization model, \texttt{nthComp}, rather than a power law, but did not find a significant difference in the fitted reflection parameters.

We also test another variation of \texttt{relxill}, called \texttt{relxillLp} which calculates reflection from a lamppost source a certain height $h$ above the compact object. In doing so, we are able to set the lamppost height above the neutron star and accretion disk, which is an important consideration in reverberation modeling (see Section~\ref{Model}). Since we model reverberation using a lamppost height of $5~R_G$, $10~R_G$, or $20~R_G$, a more fully consistent approach would be to use the lamppost version of \texttt{relxill} with these heights, fitting each height to the spectra. In doing so, we do not find a significant difference in the reflected flux fraction which we use to calculate the lags, and our results are therefore unchanged. For simplicity and in order to be more concise, we only include a discussion of results using \texttt{relxill} hereafter.

One potential complication which arises in models of a disk as the thermal component is that the best-fit disk temperatures are at or above 3 keV on average. This is much higher than typical disk systems found in atoll type NS LMXBs \citep[see, e.g.][]{Lin_etal2007}. Unrealistic temperatures are not a problem in our models including a blackbody as the thermal component: the best fit values for the blackbody temperature on average are at or above 2 keV. This value is often seen in NS LMXBs when a blackbody component is included in the model, though it is more typical of the higher luminosity Z-sources than in atoll sources like 4U~1728$-$34 \citep{Church_etal2014}.

We also attempted fits using other reflection models, including a modified version of the \texttt{reflionx} model \citep{Ross&Fabian2005}, with a variable high energy cutoff, as well as a separate modified version of \texttt{reflionx} which gives the reflected spectrum due to incident blackbody emission rather than a power law. With the variable high energy cutoff version, called \texttt{reflionx\_HC}, we find results similar to those measured using \texttt{relxill}. Reflection models using an incident blackbody spectrum could not replicate the same quality fits as our other models, or were unphysical in that the blackbody normalization providing the incident flux for reflection dropped to zero for many of the 36 spectra.

We create a representative model spectrum for each of the four QPO frequency groups (700--800, 800--900, 900--1000, and 1000--1100 Hz) by taking the weighted average for every best-fit parameter across each of the individual spectra belonging to those bins. In doing so, we combine the results of fitting all 36 \emph{RXTE} PCA spectra into the four QPO bins for which the lag-energy spectra has already been calculated by \citet{Peille_etal2015}. The resulting model values are given in Table~2.

The recent analyses of 4U~1728$-$34 by \citet{Sleator_etal2016, Mondal_etal2017} both use \texttt{relxill} to model the distinct reflection features, but include both a blackbody and a continuum Comptonized spectrum using the model \texttt{compTT}. Comparing those results with our blackbody model, the continuum blackbody temperature reported in \citet{Mondal_etal2017} ranges from 0.7 to 2.4 keV, and our values fall well within that range, albeit with lower normalizations (which is to be expected, considering their inclusion of an additional continuum component). Our measured power law index $\Gamma$ and ionization parameter $\xi$ are also consistent with those reported. While \citet{Sleator_etal2016} report an upper limit on the inner disk radius of 2.0 (in the units of the innermost stable circular orbit, or ISCO) using a variety of reflection models, \citet{Mondal_etal2017} measure the inner disk radius to be 3.1 and 3.9 ISCOs for two separate \emph{NuSTAR} observations. Such results are similar to our own reported values, seen in Table~2, though our inner disk radius measurements are generally poorly constrained and we find an increasing inner disk radius with increasing QPO frequency. This is true for our disk model where we find higher (though poorly constrained) inner disk radii for the 900--1000 and 1000--1100~Hz frequency groups. However, we note that given the limited spectral resolution of the \textit{RXTE} spectra it is hard to put significant weight on this apparent increase. Aside from these few differences, the overall agreement between our results and those reported by \citet{Sleator_etal2016, Mondal_etal2017} suggests that our spectral modeling provides a reliable approximation of the shape of the reflection spectrum of 4U~1728$-$34 during the detection of the upper kHz QPOs.

\section{Modeling the lags} \label{Model}

To model the effects of X-ray reverberation we adopt the two-dimensional (2D) relativistic ray-tracing function first calculated by \citet{Reynolds_etal1999}, as used and described by \citet{Cackett_etal2014}. The 2D transfer function tracks the disk's response to an incident impulse of X-rays in both time and energy --- the time-averaged response of the disk is the well known reflection spectrum \citep[see, e.g.][]{Ross&Fabian2005}, while the light travel time to different radii in the disk is also considered in the 2D transfer function. While the energy resolved response encodes information about the disk geometry and kinematics, the time resolved response depends primarily on the light travel-time from the source to different regions of the disk, including the general relativistic effects of the curved spacetime in the vicinity of the compact object. Though the transfer function was first calculated to model the reverberation time lags detected in AGN, X-ray reverberation has now been detected in black hole LMXBs as well \citep{Kara_etal2019, Uttley_etal2011, DeMarco_etal2015}, and time lags due to reverberation should scale roughly with the mass of the compact object \citep{Uttley_etal2014}.

Here we adapt the 2D transfer function to the neutron star source 4U~1728$-$34, following the application of the transfer function to the neutron star LMXB 4U~1608$-$52 by \citet{Cackett2016}, which tested reverberation for the lower kHz QPOs of that source. While the transfer function calculates a time lag for the Fe line produced by reflection from the disk, the upper kHz QPO lags for 4U~1728$-$34 are measured over the full (3--25 keV) energy range. Thus, our modeled lag-energy spectra must include these energies, and must take into account the full reflection spectrum in order to do so. Furthermore, the time-averaged spectrum contains both the reflected emission as well as the primary, incident emission. The incident emission observed in each band will have zero lag, and the reflected emission in each band will have zero lag --- only the incident versus reflected emission will show a lag, which has the effect of diluting lags, and this dilution of the reflected spectrum must be taken into account \citep{Kara_etal2013, Uttley_etal2014, Cackett_etal2014, Cackett2016}. In order to adjust the transfer function for these dilution effects and to model the response at all energies, we calculate a reflected flux fraction across the energy spectrum from our spectral models. For each model and in each QPO bin, we obtain the reflected flux fraction at a given energy by dividing the unblurred, reflected flux by the total flux at that energy. The reflected flux fraction for our two models is given for each QPO bin in Figure~\ref{refl_frac}. Since illumination of the accretion disk should produce thermal emission by the disk itself \citep{Ross&Fabian2007}, we include the disk component as part of the reflected spectrum in the disk model. This thermal reverberation is what dominates the shape of the reflected flux fraction shown in the left of Figure~\ref{refl_frac}. Whether or not thermal reverberation is included in the reflected flux is the primary difference between the disk and blackbody models, rather than the choice of thermal component used in spectral fitting.

\begin{figure}[h]
\includegraphics[trim=2.0cm 5.57cm 2.5cm 4.5cm, clip=true, width=9cm]{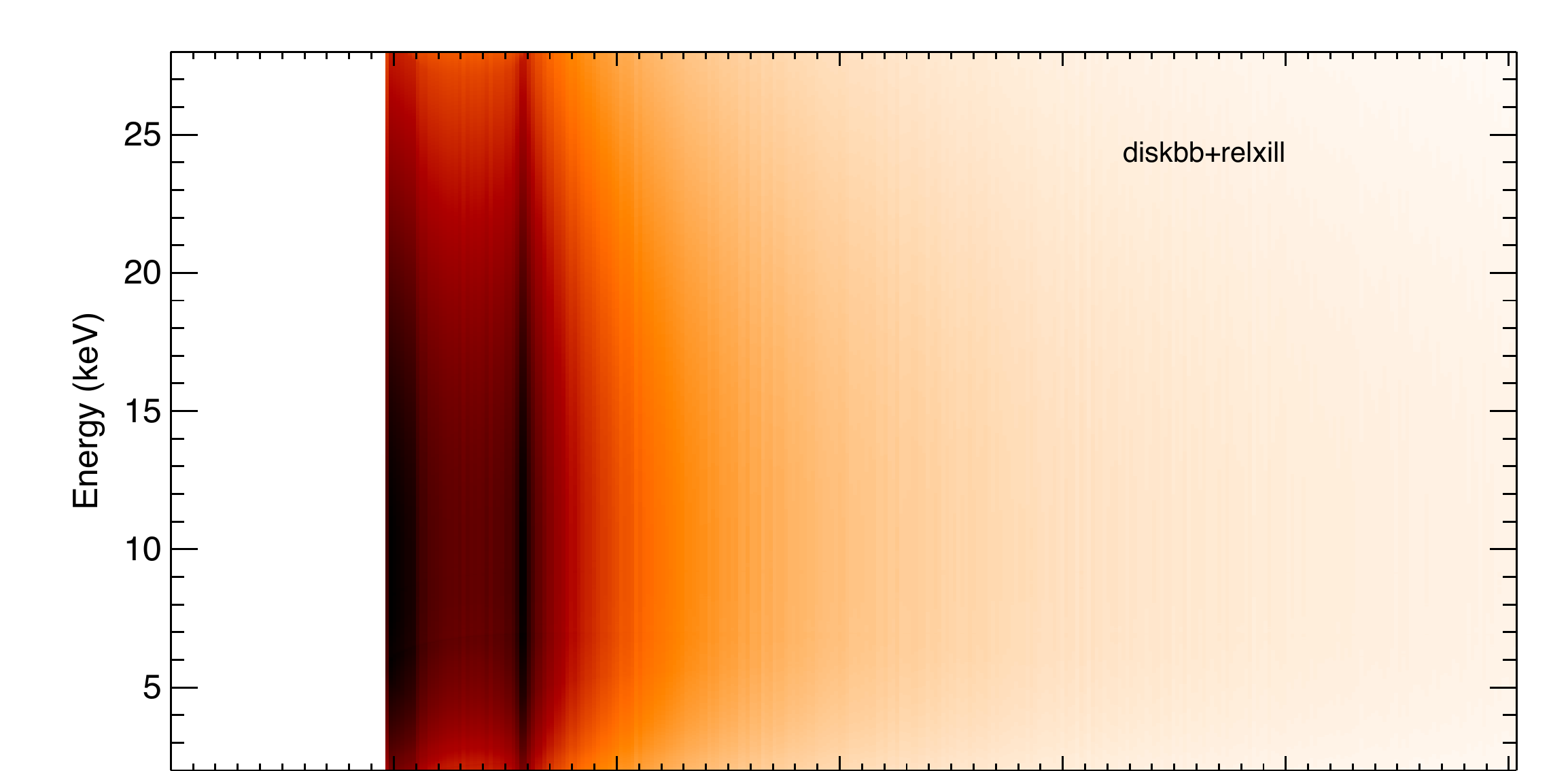}
\includegraphics[trim=2.0cm 5.57cm 2.5cm 5.24cm, clip=true, width=9cm]{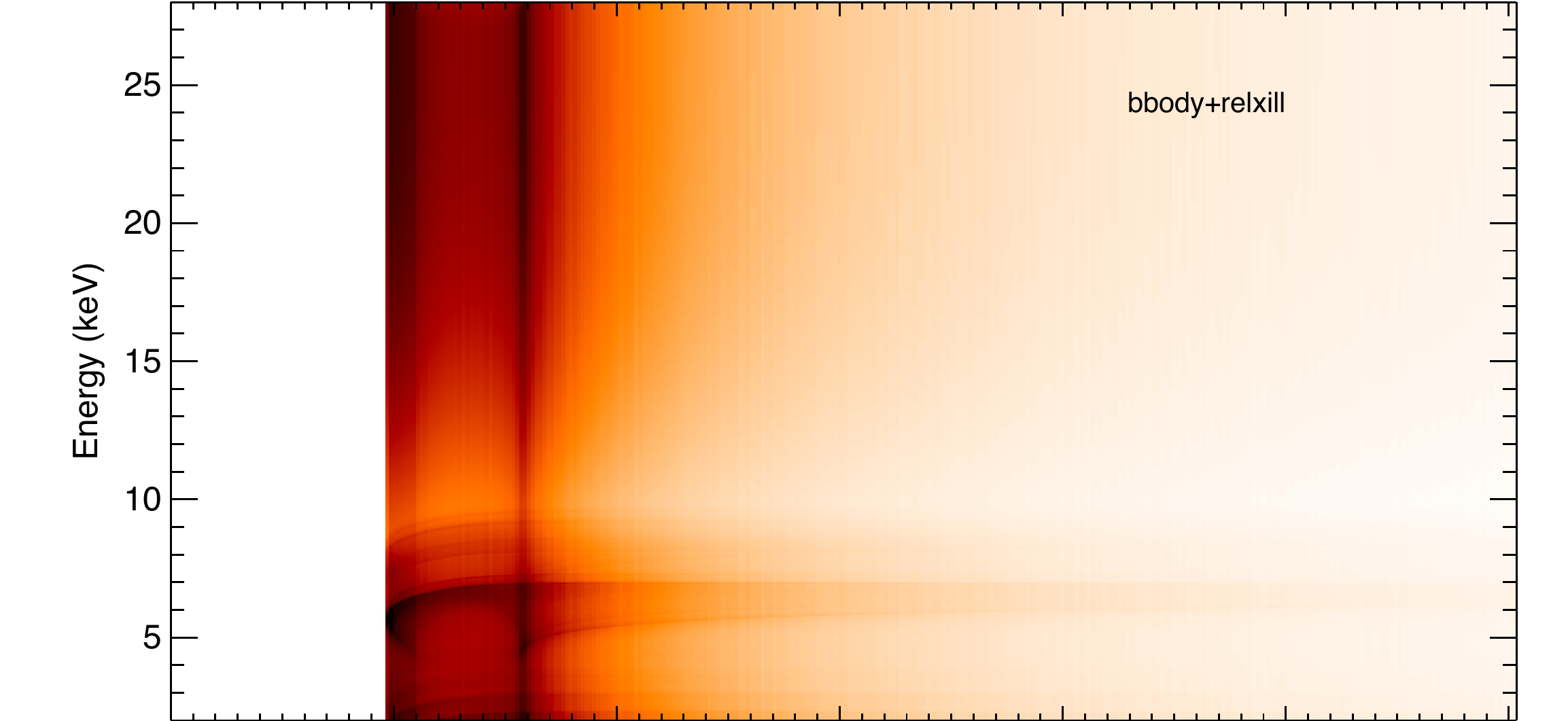}
\includegraphics[trim=2.0cm 4.0cm 2.5cm 5.24cm, clip=true, width=9cm]{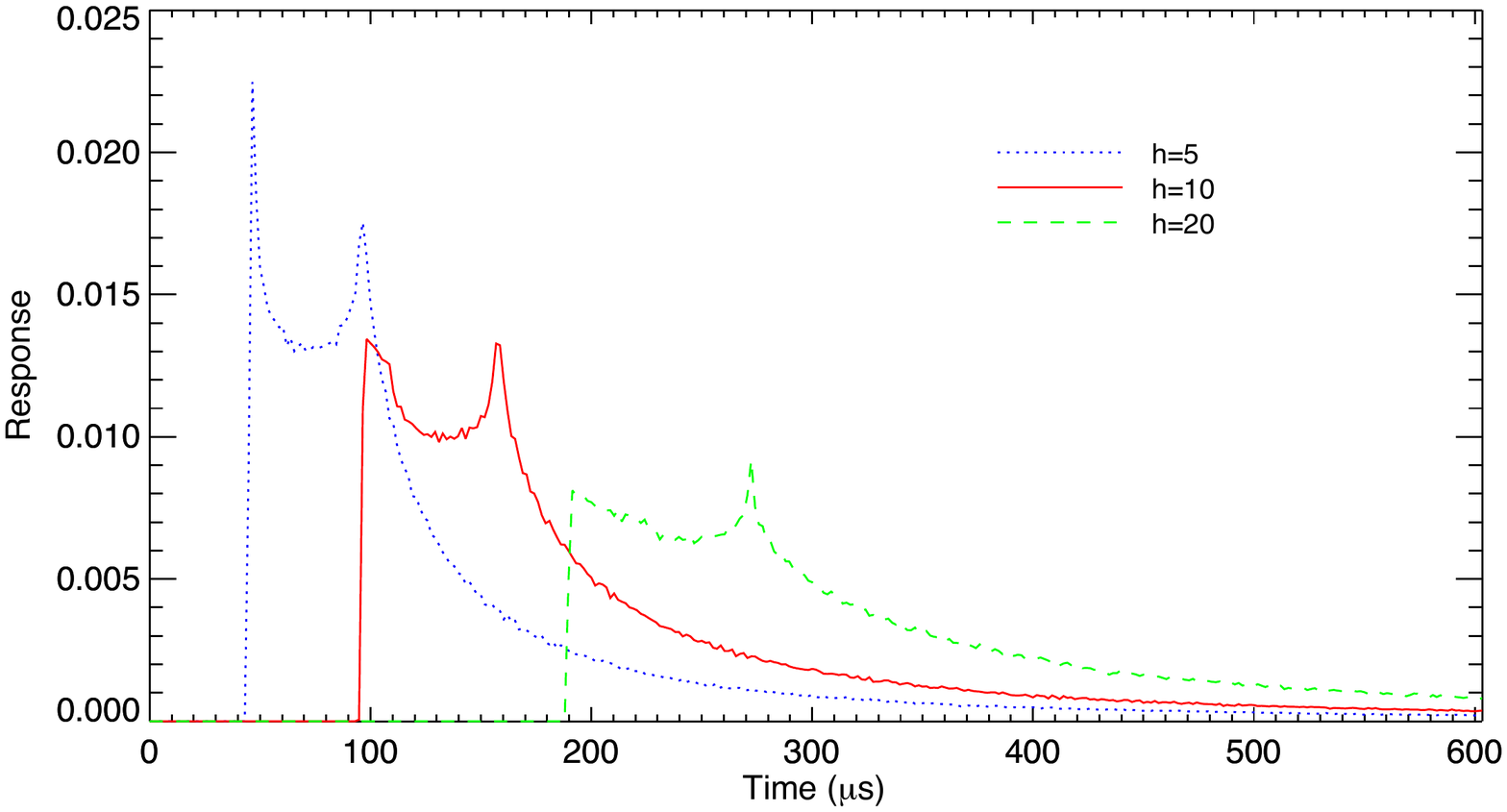}
\caption{Dilution-corrected two-dimensional transfer function for the 700--800 Hz QPO bin in 4U~1728$-$34, shown for both the disk model (upper) and blackbody model (middle) and for a source height of $h=10~R_g$. The transfer function encodes the disk's response to a delta function flare in intensity at a given height above the disk, both in time and energy, and is the convolution of the reflected fraction from Figure~\ref{refl_frac} with the timing response at the Fe line energy. The timing response, averaged over all energies, is also shown for emission from a height of 5, 10, and 20 $R_G$ above the disk.}
\label{2dTF}
\end{figure}

Using the method of \citet{Cackett2016}, we convolve our reflected flux fraction with the calculated 2D transfer function for 4U~1728$-$34. We use transfer functions that assume a compact object mass of 1.4 $M_\odot$, an inclination of 32 degrees, a dimensionless spin parameter of $a = 0.15$, and a source height of 5, 10, or 20 $R_G$ above the disk. This produces the dilution corrected transfer function specific to 4U~1728$-$34 when upper kHz QPOs were detected, which is given in Figure~\ref{2dTF}, along with the timing response of the disk for each source height. Once the disk's response in time and energy is computed for the entire energy band, lags can be calculated between different energy bins to produce a lag-energy spectrum.

The dilution-corrected transfer function is arbitrarily normalized at 6.4 keV, following \citet{Cackett_etal2014} and \citet{Cackett2016}, and the lag is calculated as the phase of the Fourier transform of the timing response of the disk. We use a reflected response fraction of 0.5 for the initial normalization, though changes in this simply scale the amplitude of the lags without changing their shape \citep{Cackett_etal2014}. This is done using the models fit to each QPO frequency group, and the lag is calculated by summing over the Fourier frequencies within the binning of that QPO frequency group (i.e., summing over 700--800 Hz). The lag is then converted into a physical time (in microseconds) assuming a canonical 1.4 $M_\odot$ neutron star, since time lags due to reverberation scale with mass.

\begin{figure}[t]
\includegraphics[trim=2.5cm 5.5cm 0.0cm 5.0cm, clip=true, width=10cm]{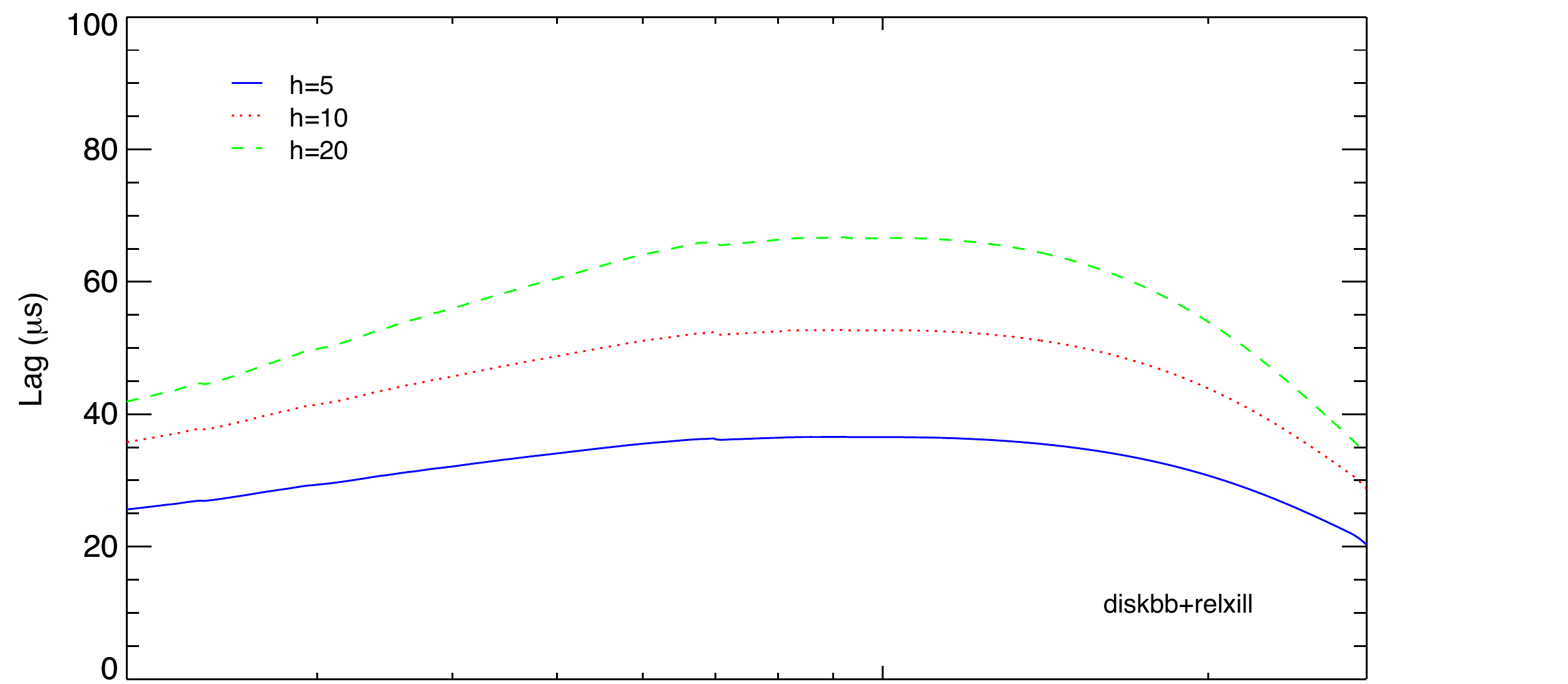}
\includegraphics[trim=2.5cm 4.0cm 0.0cm 5.2cm, clip=true, width=10cm]{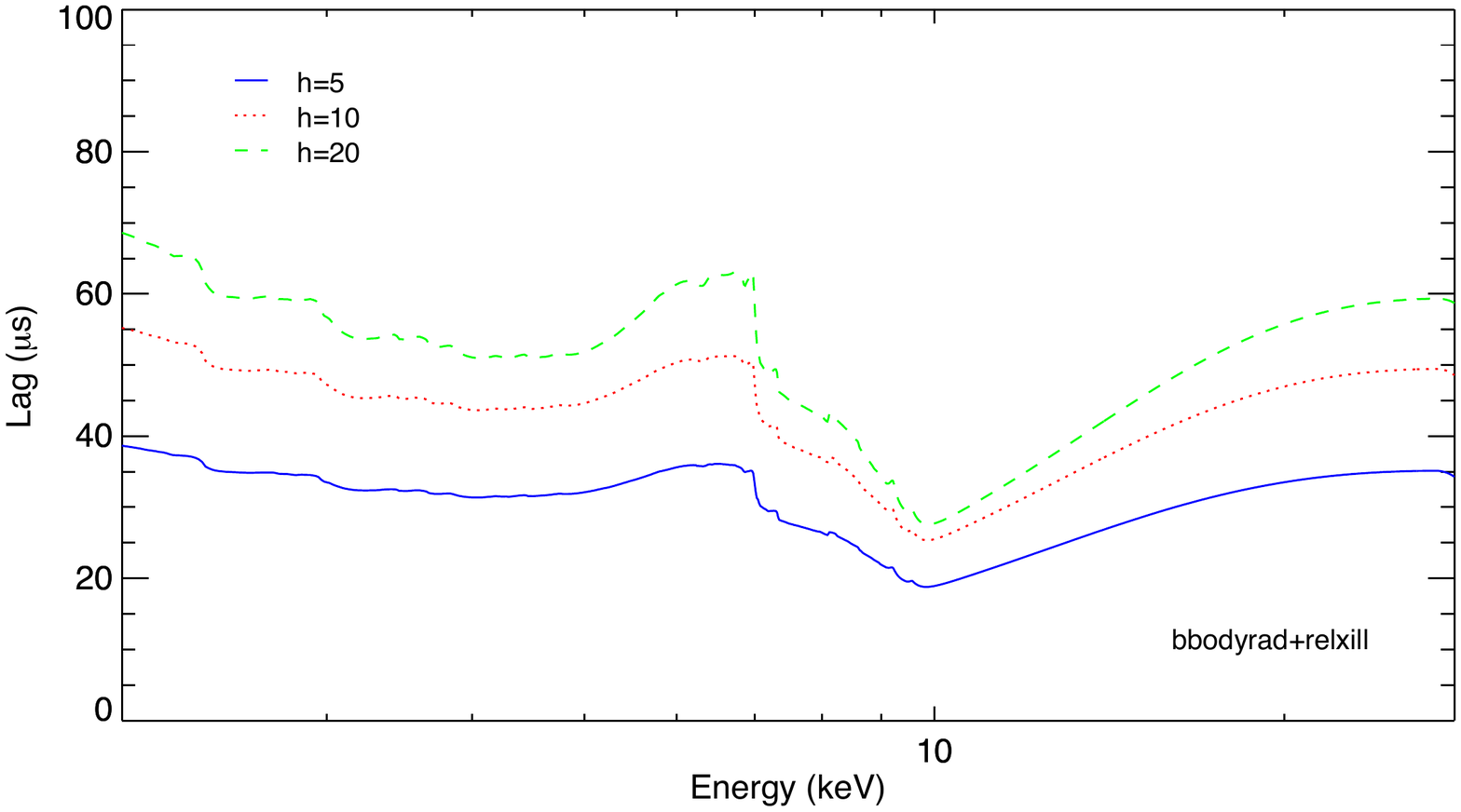}
\caption{The lag-energy spectrum of two different spectral models, for the 700--800 Hz QPO bin in 4U~1728$-$34. The disk model (upper) includes thermal reverberation, and shows a smooth, broad curve, which is either flat or slightly curved with a peak at higher energies, depending on the QPO frequency group modeled. The blackbody model (lower) provides a distinctly different lag-energy spectrum, with strong reflection features at $\sim$ 6--7 keV, and a minimum at $\sim$ 9 keV before increasing sharply at higher energies. The lag-energy spectra using three different source heights are shown for both models, with the dashed green line representing a source height of 20 $R_G$, the dotted red line using h $= 10 R_G$, and the solid blue line using a source height of just 5 $R_G$ above the disk and neutron star.}
\label{lag_model}
\end{figure}

The lag-energy spectrum is shown for both the disk and blackbody models in Figure~\ref{lag_model} for the 700--800 Hz QPO frequency group. As can be seen, the choice of continuum thermal component has a drastic effect on the resulting lag energy spectrum. This is almost entirely due to the inclusion of thermal reverberation in the disk model, which has the effect of diluting the key features of the reflection spectrum (i.e., the broadened Fe line and the Compton hump). These features are fairly obvious in the blackbody model, which does not include thermal reverberation. We test whether just a fraction of the thermal disk emission should be included, and find that reflection features become stronger and that the general shape resembles the results using a single-temperature blackbody as the fraction of the disk representing thermal reverberation decreases (see the discussion in Section~\ref{Discuss}). The strong reflection features of the blackbody model provide a hard lag at higher energies, mainly due to the broad Compton hump at those energies and the lack of dilution from thermal reverberation, which makes the blackbody model more promising when considering the measured lag-energy spectra of \citet{Peille_etal2015}. The same qualitative differences between models is seen in all four QPO bins, and for the different versions of \texttt{relxill} as well.

\begin{figure*}[t!]
\includegraphics[trim=4.0cm 5.55cm 6cm 5.25cm, clip, width=9cm]{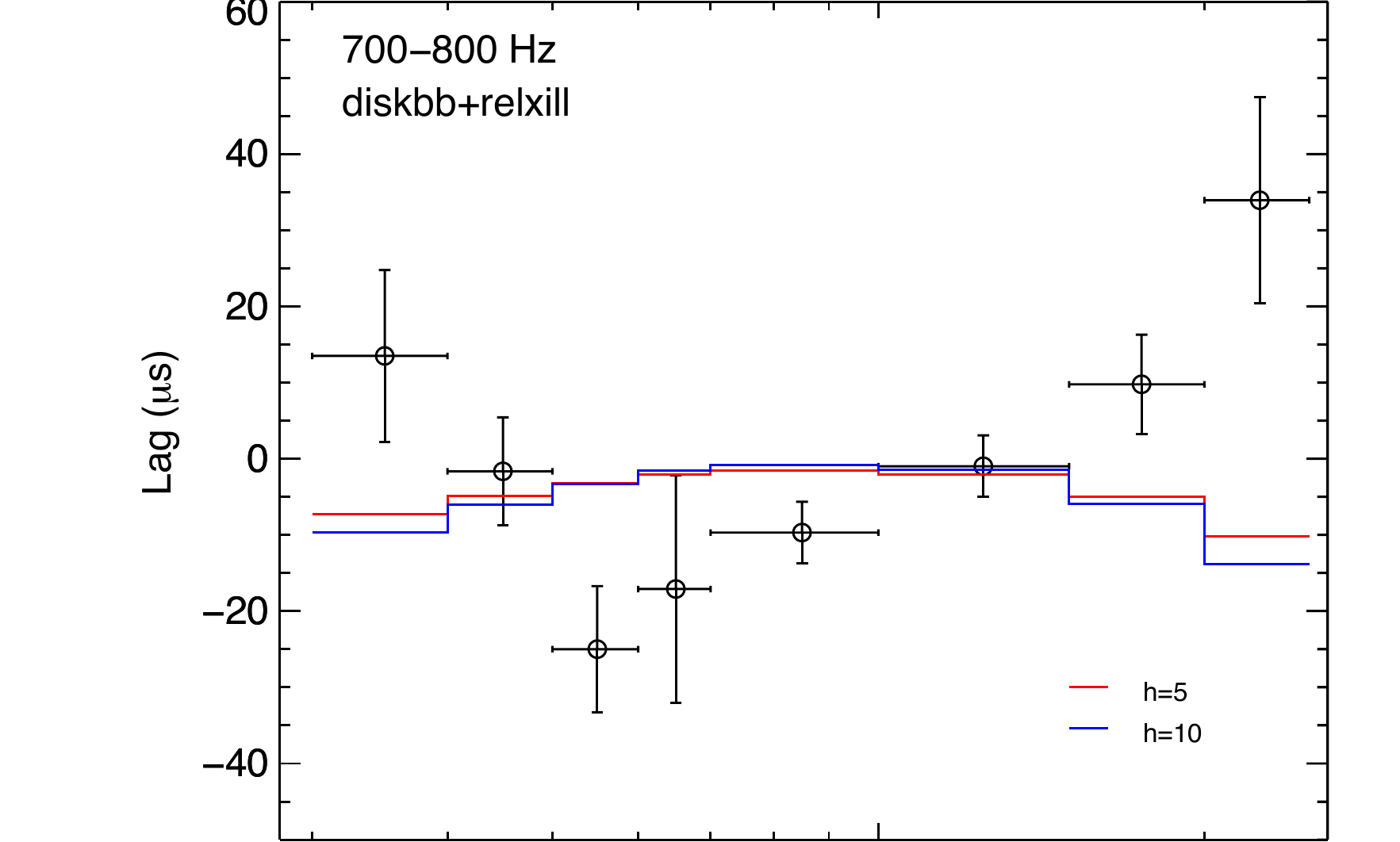}\includegraphics[trim=7.5cm 5.55cm 2.5cm 5.25cm, clip, width=9cm]{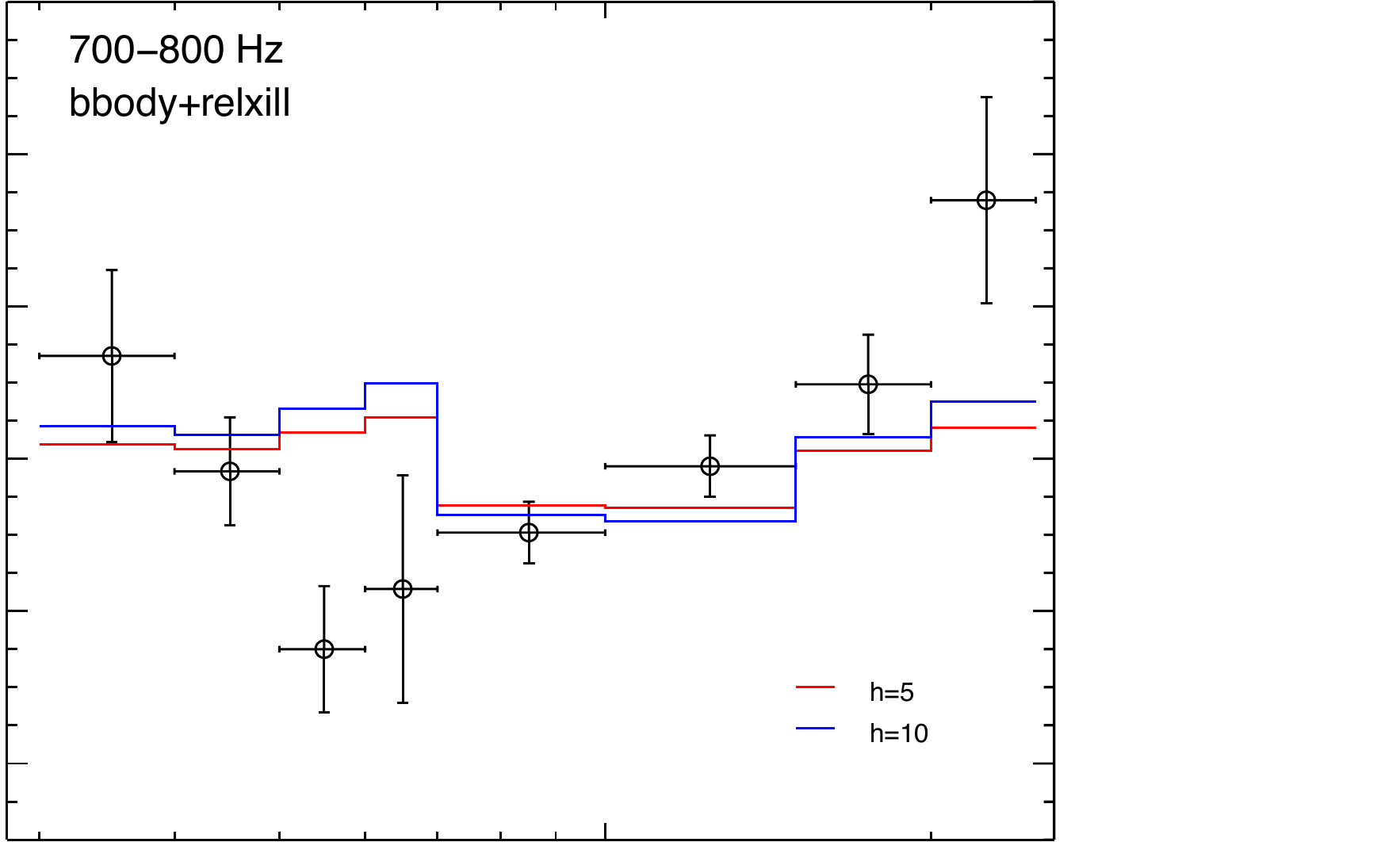}
\includegraphics[trim=4.0cm 5.55cm 6cm 5.25cm, clip, width=9cm]{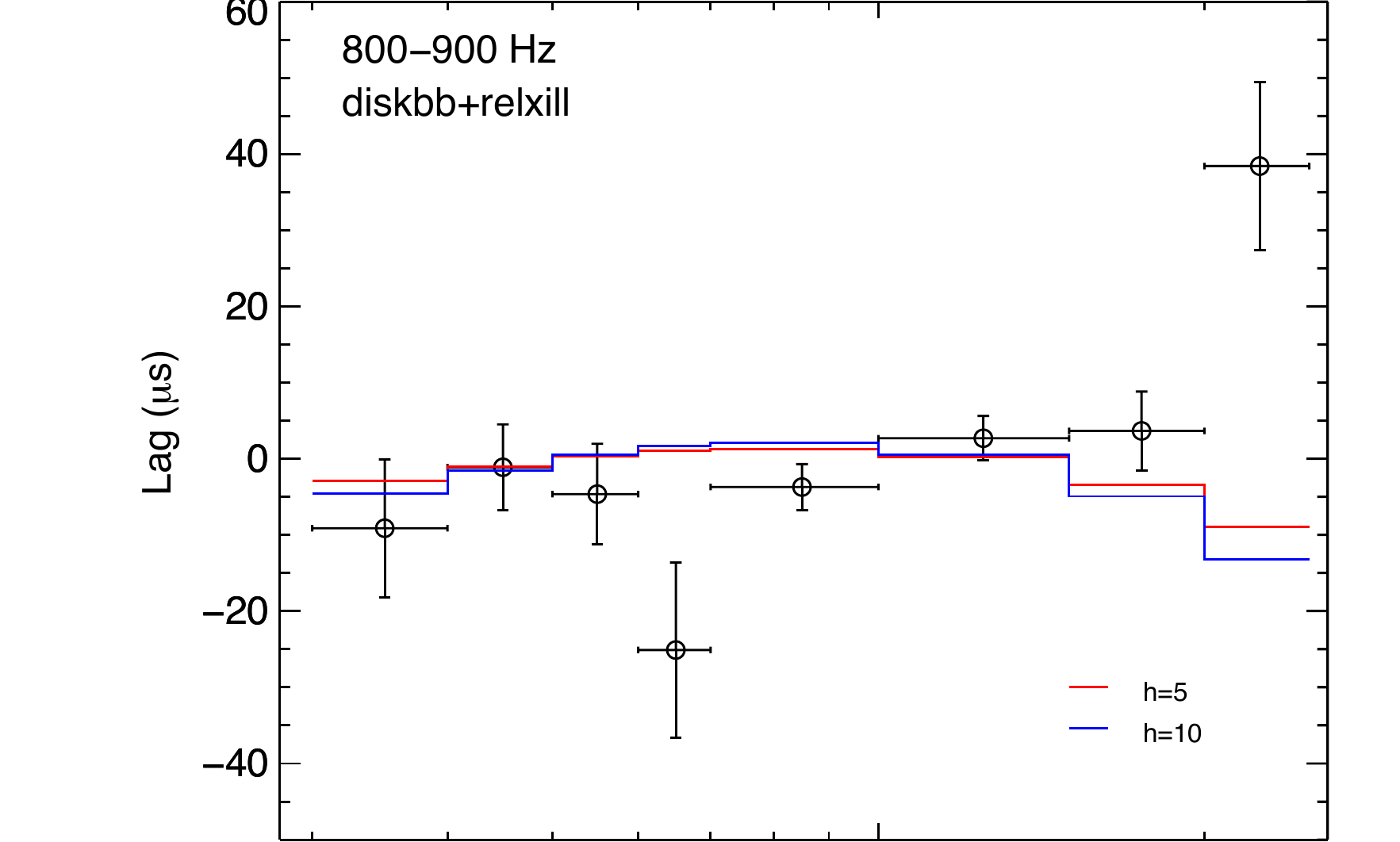}\includegraphics[trim=7.5cm 5.55cm 2.5cm 5.25cm, clip, width=9cm]{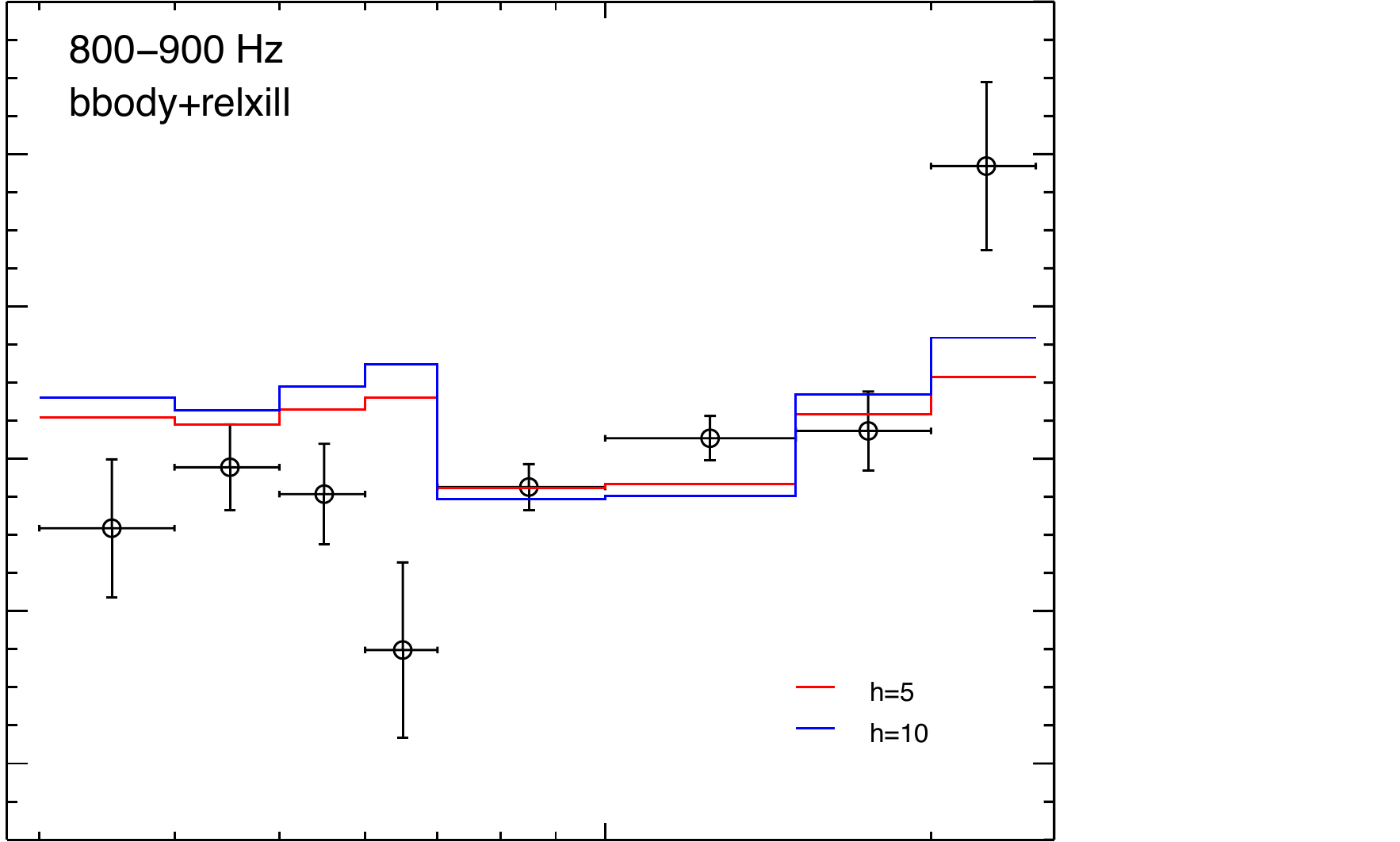}
\includegraphics[trim=4.0cm 5.55cm 6cm 5.25cm, clip, width=9cm]{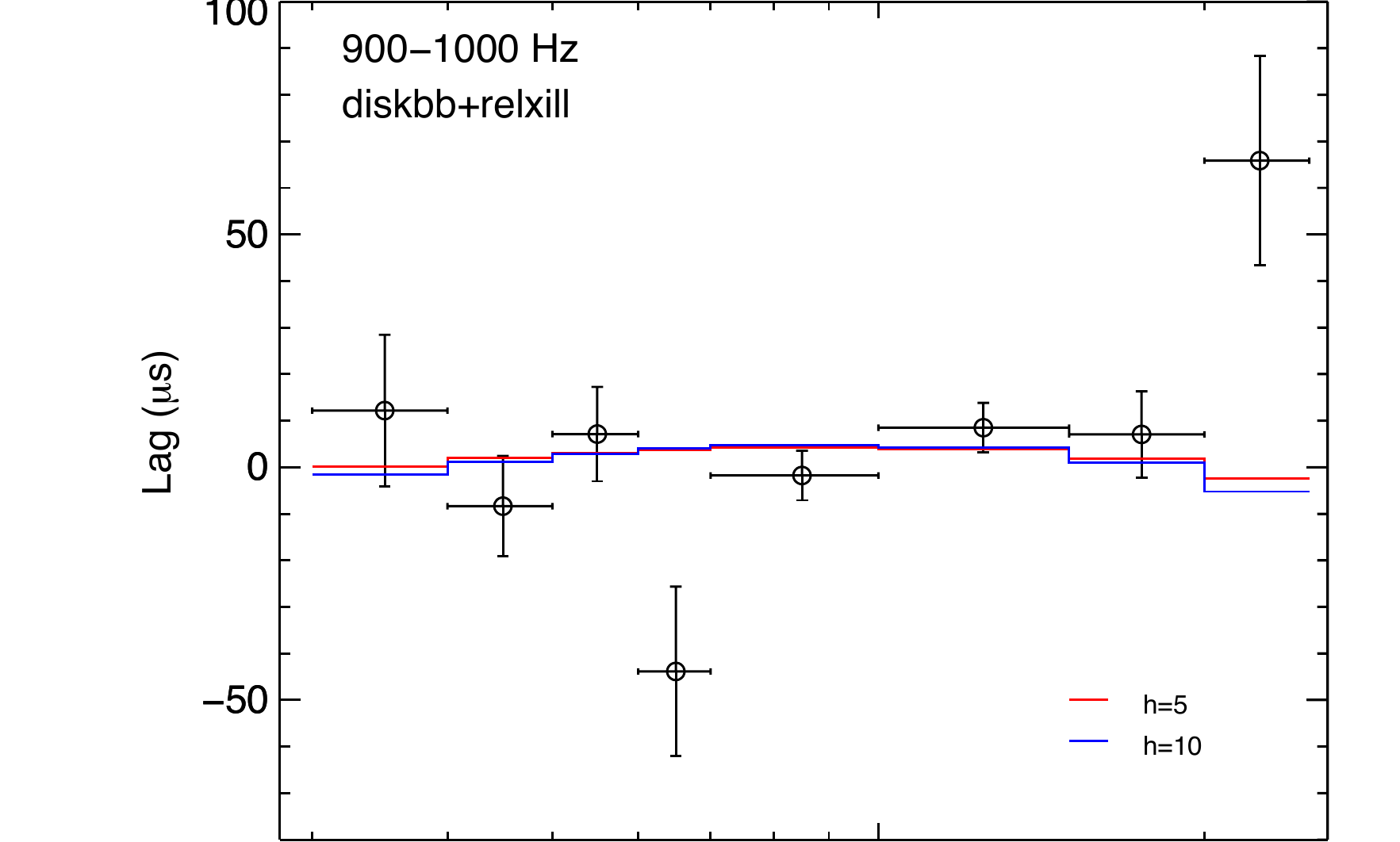}\includegraphics[trim=7.5cm 5.55cm 2.5cm 5.25cm, clip, width=9cm]{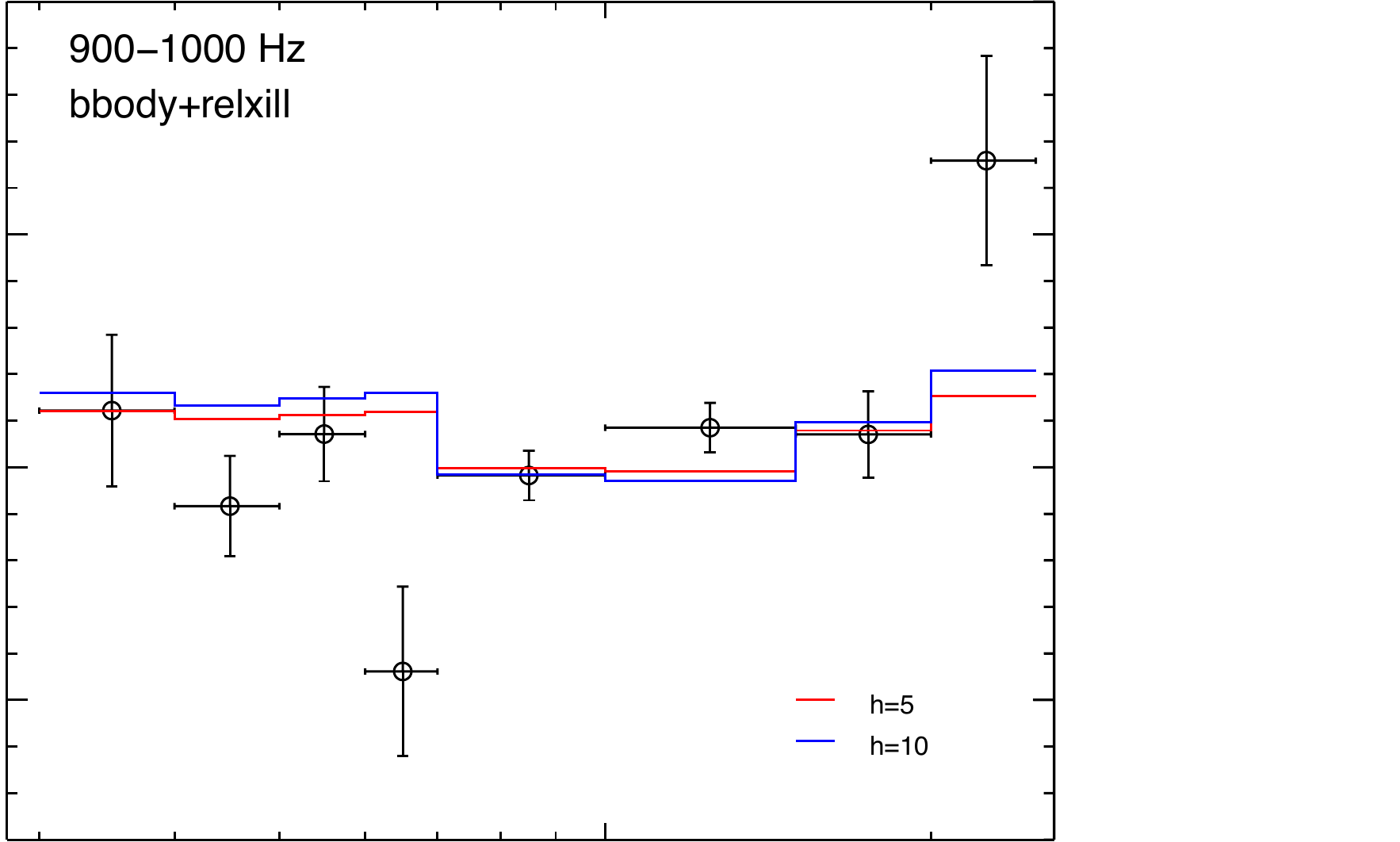}
\includegraphics[trim=4.0cm 4.0cm 6cm 5.25cm, clip, width=9cm]{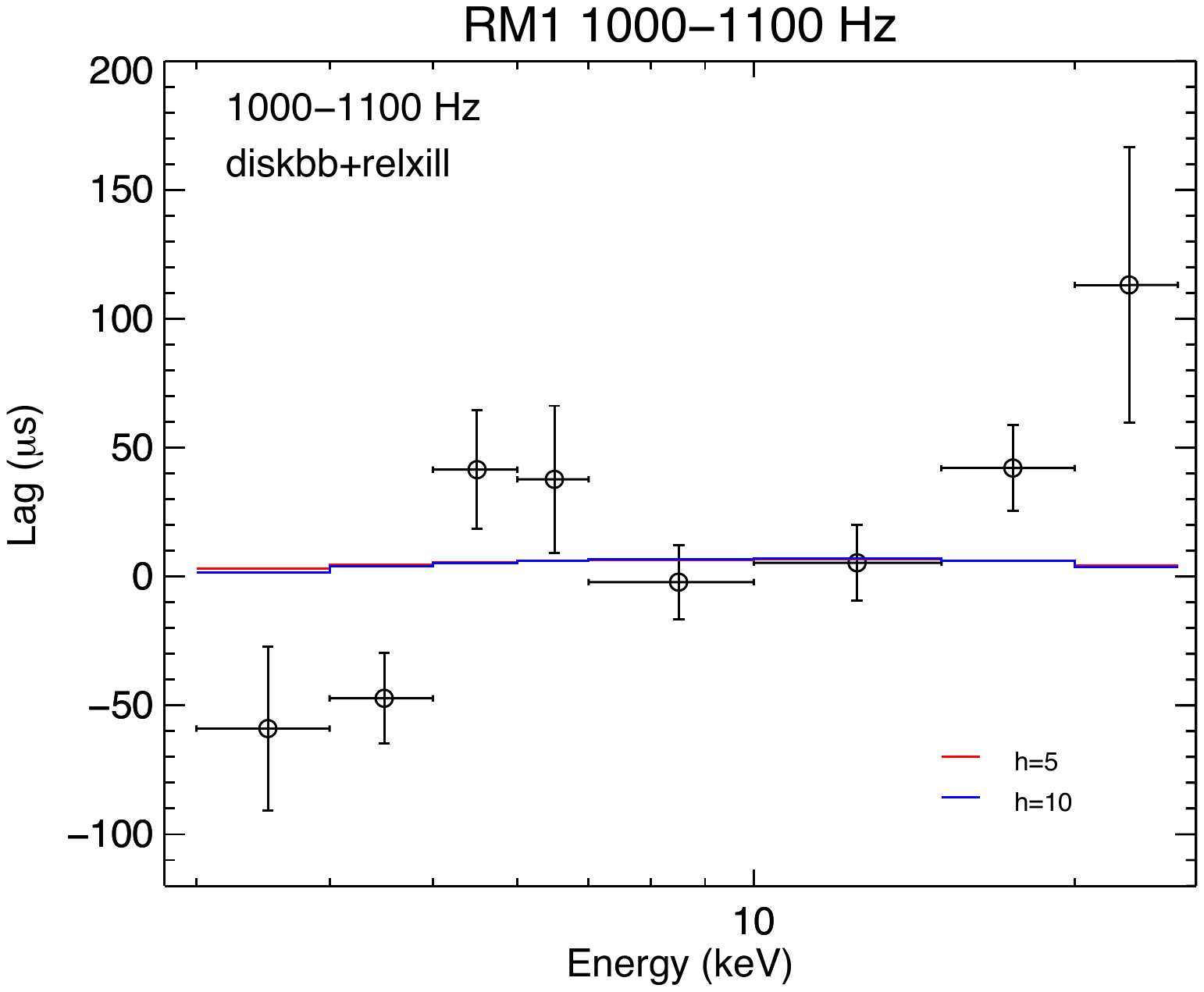}\includegraphics[trim=7.5cm 4.0cm 2.5cm 5.25cm, clip, width=9cm]{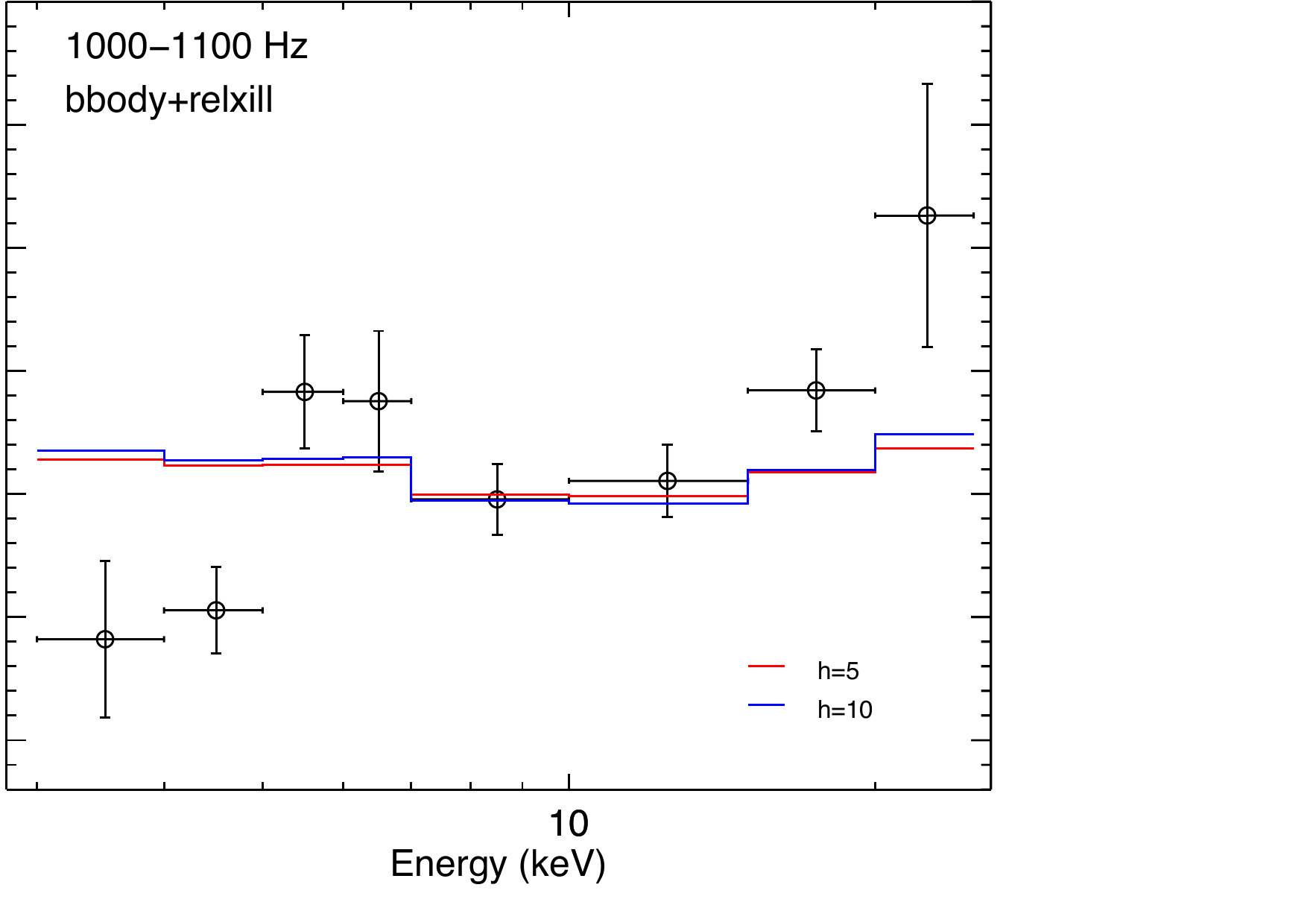}
\caption{Lag-energy model spectra (see, e.g., Figure~\ref{lag_model}) are binned and shifted to fit the data as measured by \citet{Peille_etal2015}. Both models are shown, and both provide extremely poor fits to the data. The disk models (left) are flat and do not provide the hard lags at higher energies seen in the data, while the blackbody models do not match the shape of lags below $\sim$ 10 keV, despite having a somewhat promising hard lag above 10 keV.}
\label{lag_data}
\end{figure*}

At first glance, the increase in lags at high energies in the blackbody model (see Figure~\ref{lag_model}, lower) appears similar to the hard lags measured for 4U~1728$-$34 \citep{Peille_etal2015, Troyer_etal2018}. We fit our model lag-energy spectrum for each QPO frequency group to the lags measured by \citep{Peille_etal2015}, by shifting the lags up or down with the addition (or subtraction) of a constant. Since the lags measured are relative lags --- in each energy bin the lag was calculated against the 3--25 keV reference band minus that energy bin --- we are free to shift our model along the lag axis without changing the relative lag between different frequencies. As shown separately for each of the four QPO frequency groups in Figure~\ref{lag_data}, the shape of our models is inconsistent with the measured lag-energy spectra. Fit results for the shifted models are given in Table~3. With the disk model, the convex shape of the spectrum is incompatible with the concave measured lags, particularly at higher energies where the lags should increase. The blackbody model shows a strong, broad Fe line component, which is not seen in the data for the lowest three frequency groups. While a feature between 5 and 7 keV does appear in the 1000--1100 Hz lag-energy spectrum, it should be stressed that this QPO bin has particularly poor statistics, and the visible bump may not be highly significant. In any case, our blackbody model for the 1000--1100~Hz frequency group shows the weakest Fe line component relative to the other frequency groups, which is the opposite trend seen in the data, and our model for this frequency group still provides a poor fit to the data. Furthermore, the measured increase at high energies in the data is greater than that seen in our models.

\begin{deluxetable*}{llllll}[t]
\tablecolumns{6}
\tablecaption{Fit Statistic for Lag-Energy Models}
\tablehead{Source Height & Spectral Model & 700--800 Hz & 800--900 Hz & 900--1000 Hz  &  1000--1100 Hz}
\startdata
\multirow{2}{*}{$h = 5~R_G$} & \texttt{disk+relxill} & 31.4 & 30.0 & 20.1 & 25.3 \\
& \texttt{bbody+relxill} & 24.6 & 25.4 & 20.8 & 26.1 \\
\tableline
\multirow{2}{*}{$h = 10~R_G$} &\texttt{disk+relxill} & 35.7 & 35.3 & 21.1 & 25.0 \\
& \texttt{bbody+relxill} & 27.4 & 32.5 & 24.2 & 26.7 \\
\tableline
\multirow{2}{*}{$h = 20~R_G$} &\texttt{disk+relxill} & 41.8 & 41.6 & 22.1 & 24.7 \\
& \texttt{bbody+relxill} & 32.1 & 40.3 & 26.4 & 26.8 \\
\tableline
\multirow{2}{*}{$h = 60~R_G$} &\texttt{disk+relxill} & 19.3 & 15.5 & 16.6 & 26.7 \\
& \texttt{bbody+relxill} & 28.2 & 29.3 & 20.0 & 26.6 \\
\enddata
\tablecomments{$\chi^2$ values demonstrating the poor statistical fit of our lag-energy models to the data, for each source height and QPO frequency group. We add or subtract a constant to each of our lag-energy models to allow direct comparison to the data of \citet{Peille_etal2015}, where lags were measured relative to a reference band. In each case, we have just 7 degrees of freedom, giving a reduced $\chi^2$ of $>$ 2 in even the best cases.}
\end{deluxetable*}

We notice that the shapes of the lags from reverberation do not match those of the measured lags. In fact, we find that our models produce almost the opposite trend, at least when including disk reverberation, which produces a convex or `n' shape, while the lags follow a more concave `v' shape. Our blackbody models, without thermal reverberation, do show a general `v' shape with a promising hard lag at higher energies, yet they also show a strong Fe line hump around 6--7 keV. This is exactly the opposite of the slight dip seen in the measured lags at 6--7 keV for the lower three QPO frequency groups (excluding the 1000--1100 Hz group).

This discrepancy might be resolved if we allow for phase wrapping of the lags \citep[for a discussion of these effects, please see][]{Uttley_etal2014, Cackett_etal2014}. Phase wrapping results in negative lags at high enough frequencies, and a negative lag would flip the shape of the entire lag-energy spectrum. This might provide a better match to the measured lags of 4U~1728$-$34.

\begin{figure}[h!]
\includegraphics[trim=4cm 4cm 2cm 5cm, clip=true, width=10cm]{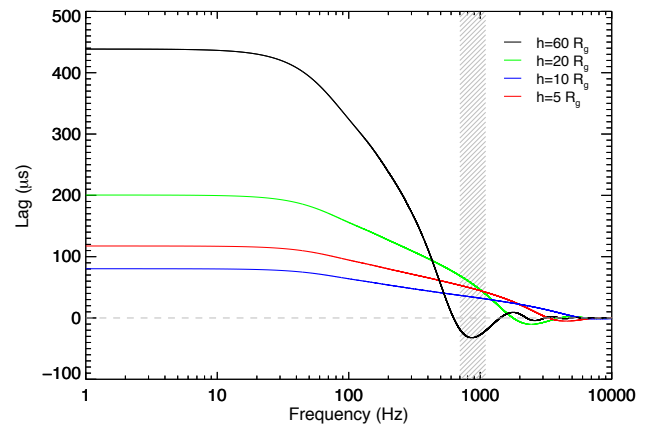}
\caption{Magnitude of the lags plotted against frequency for our reverberation model with various source heights. The lags increase with source height, while the frequencies at which phase-wrapping becomes important decrease with source height. The frequency span of the upper kHz QPOs we consider are highlighted in grey, from 700--1100 Hz. As can be seen, with h=60 $R_g$ (black), the lags are phase wrapped and negative for the frequencies of the upper kHz QPOs. Lags are calculated assuming a canonical 1.4 $M_\odot$ neutron star.}
\label{lag_freq}
\end{figure}

We are constrained to model the 700--1100 Hz frequency range of the detected upper kHz QPOs for 4U~1728$-$34. It is possible for phase-wrapping to occur at these frequencies if we allow for an increase in the total time delay, due to either a more massive compact object or an increased height of the irradiating source above the disk. In the case of a more massive neutron star, since the magnitude of the lags scales directly with the mass of the compact object, it is a straightforward calculation to find the frequencies at which phase wrapping will begin to occur. Following the dependence of the lags on black hole mass from \citet{Cackett_etal2014}, we find that even for phase wrapping to begin at 1200 Hz with a source height of 10 $R_G$, the compact object must have a mass of $\sim$~2.9 M$_\odot$ --- far too heavy for a typical neutron star, and so we neglect this possibility. Increasing the height of the hard X-ray source above the disk will also naturally produce longer time delays, and in the case of a 1.4 $M_\odot$ neutron star and a source height of 60 $R_G$, the frequencies at which phase wrapping will occur lines up with the measured frequencies of the upper kHz QPOs, as can be seen in Figure~\ref{lag_freq}.

We therefore model the lags with h = 60 $R_G$, with and without disk reverberation. In the case of the disk models, where thermal reverberation is included, the flipped lag-energy spectrum provides a hard lag at higher energies and more generally matches the `v' shape of the measured lags. Due to the change in shape, we are able to find our best fits yet to the lag data, however, the fit statistics are still very poor (as can be seen in Table 3). In particular, our models do not fit the dip between 5--8 keV and are too flat at higher energies. In the case of the blackbody models (without thermal reverberation), the phase wrapped lags are generally able to fit the lower energies quite well, but are no longer able to fit the hard lags at higher energies and are not an improvement on our previous results.

\subsection{Testing an Additional Reverberation Model}

Our method for calculating the reverberation signature from the full reflection spectrum makes some simplifications and assumptions. For instance, our impulse response function is calculated for a single emission line only, and does not use the full reflection spectrum from each location on the disk. Instead, to approximate this we convolve the impulse response function with the reflection flux fraction at each energy from the best-fitting spectral models. A comprehensive approach is to calculate the time lag and full reflection spectrum at each location on the disk \citep[e.g.,][]{Wilkins_etal2013, Wilkins_etal2016, Chainakun_etal2015, Chainakun_etal2016, Mastroserio_etal2018, Mastroserio_etal2019, Ingram_etal2019}.

In order to verify that the simplifications in our approach does not cause a large effect, we compare our results with the recently developed, more self-consistent, public reverberation model called \texttt{reltrans}\footnote{\url{https://adingram.bitbucket.io/}} \citep{Mastroserio_etal2018, Mastroserio_etal2019, Ingram_etal2019}. The \texttt{reltrans} model calculates the disk's response to variations in the incident power law or Comptonized flux, which are represented by changes in the index or normalization of the power law. These variations may correspond to physical changes in the accretion rate or temperature of the Comptonizing region. The model self-consistently calculates the reflection fraction, and takes into account phase lags at all frequencies along with the variability amplitude at all energies \citep{Mastroserio_etal2018, Mastroserio_etal2019, Ingram_etal2019}. For the rest-frame reflection spectrum, \texttt{reltrans} uses the model \texttt{xillver} (as we also use, considering \texttt{xillver} is the rest-frame spectrum on which \texttt{relxill} is based), and the time-averaged \texttt{reltrans} spectrum is identical to the version of \texttt{relxill} which also utilizes a lamppost geometry, known as \texttt{relxillLp}. We create a representative lag-energy spectrum with \texttt{reltrans} by using the model parameters established from our spectral fitting, outlined in Section~\ref{Spectra}, and find that the resulting lag-energy spectra share the same shape as our blackbody models, and that the new \texttt{reltrans} lags provide a similarly poor statistical representation of the measured lag data. \texttt{reltrans} can also be used to include hard continuum lags in addition to the reverberation lags. We find that including a hard lag still does not improve the fits. Regardless of the model, the shape of the lag-energy spectrum due to reverberation is inconsistent with the measured lags for 4U~1728$-$34.

\subsection{Testing the Variability of Model Components}

Our modeling so far uses the time-averaged spectral fits to determine the reflected flux fraction as a function of energy. However, it is the shape of the reflected flux fraction from the \emph{variable} spectrum that is important for determining the lags. Our approach assumes that the time-averaged and variable spectrum have the same shape, or at least that the variability in each model component has the same dependence on energy. It is well known that the fractional RMS amplitude for both kHz QPOs increases with energy between 3 and 10 keV \citep[see, e.g.][]{Berger_etal1996, Zhang_etal1996, Mendez_etal2001, Gilfanov_etal2003, Troyer_etal2018}.

We therefore investigate another approach, and try to work backwards from the observed lag-energy spectrum to determine if there is any combination of the best-fitting spectral components that would provide a good fit to the observed lags. We require that the variable spectrum consist of the same shape spectral components, but with each component allowed to contribute a different fraction towards the variable flux. For each QPO frequency group, we create a reflected flux fraction that best represents the shape of the measured lags. We then fit the representative reflected flux fraction using our spectral models allowing each component to contribute a different fraction of the variable flux.

\begin{figure}[t!]
\includegraphics[trim=5.5cm 4.2cm 3cm 5cm, clip=true, width=10cm]{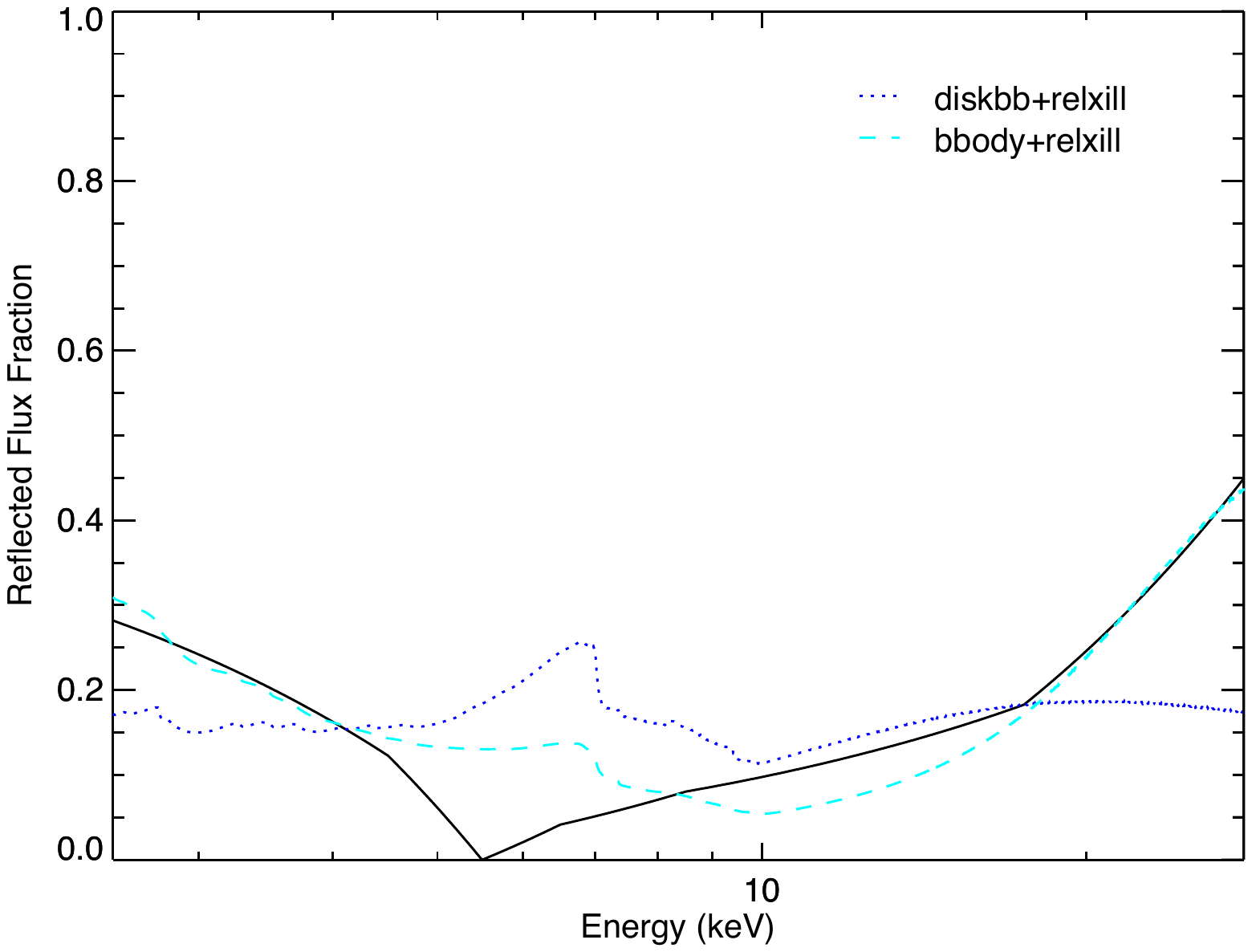}
\caption{Best-fit results for the 700--800 Hz frequency group reflected flux fraction, when testing the variability of model components. The representative reflected flux fraction (solid black line) was calculated to be able to reproduce the lag data. We plot our best-fit results allowing for each model component to vary, with the disk model shown in blue (dotted line) and the blackbody model shown in cyan (dashed line). Neither provides a good fit to the representative reflected flux fraction, and therefore neither will be able to reproduce the measured lag-energy spectrum.}
\label{back_fit}
\end{figure}

Doing this, we find that our best-fit models cannot fit the shape of the representative reflected flux fraction, and will therefore not be able to produce the correct shape of the lag-energy spectra even allowing for different variability amplitudes for each component. Our best-fit results for the 700--800 Hz frequency group are shown in Figure~\ref{back_fit}. While we are able to fit the flatter part of the reflected flux fraction between 10 and 20 keV, significant deviations occur above 20~keV or in the 5--8~keV region of the Fe line. Above 20 keV, the data rises sharply for each of the four QPO frequency groups (see Figure~\ref{lag_data}), while our best-fit models remain flat. In the 5--8~keV Fe line region, our models show a positive bump in the reflection fraction due to the Fe line, while the lag-energy data for all of the QPO frequency groups (except the 1000--1100~Hz bin) show a drop in that energy range. Thus, for the 700--800 Hz, 800--900 Hz, and 900--1000 Hz frequency groups, our models deviate dramatically from the shape of the reflected flux fraction inferred from the data between 5 and 8 keV. In the 1000--1100 Hz group, because of the bump in the lag-energy spectrum at these energies, our models fit this energy range comparatively well when we scale the variability of each model component. At the same time, however, our models for this frequency group provide an extremely poor fit below 5 keV and above 18 keV. Furthermore, it should be stressed that for the upper kHz QPOs above 1000 Hz the statistics are much worse given the lower number of total counts available. While the bump between 5--8 keV appears clearly in the lag-energy spectrum, whether or not it is highly significant or indicates the presence of an Fe line in the lags remains unclear.

In summary, it is apparent that even when allowing the normalization of the variability of each model component to vary, reverberation is unable to explain the lags of the upper kHz QPOs.

\section{Discussion} \label{Discuss}
 
While it has already been shown that the lag-energy spectrum of the lower kHz QPOs cannot be explained by reverberation \citep{Cackett2016}, such a test had not been carried out for the upper kHz QPOs. In general, the upper kHz QPOs show markedly different spectral-timing features than the lower kHz QPOs and, in sources with the best photon statistics for the upper kHz QPOs (4U~$1728-34$ and 4U~$1636-53$), the lag-energy spectra show a relatively flat lag which then increases at higher energies \citep{Troyer_etal2018}. Furthermore, the possibility of an Fe line feature in the lag-energy spectrum for the highest frequency QPOs in 4U$1728-34$ suggests that X-ray reverberation may produce the measured lags \citep{Peille_etal2015}. 

We modeled the time-averaged spectra of 4U~1728$-$34 where upper kHz QPOs were significantly detected by \citet{Peille_etal2015} with a variety of spectral models. Early tests with simple phenomenological models resulted in poor fits to the spectra and showed a broad Fe line in the residuals, suggesting a reflection spectrum. Furthermore, reflection is required in order to model reverberation, and so we tested reflection models with different sources of irradiating flux --- that could be a power law representing Comptonization (as in the case of \texttt{relxill}), a more physical Comptonization model like \texttt{relxillCp}, or a thermal source of incident flux (e.g. \texttt{reflionx$_\texttt{bb}$}). As discussed in Section~\ref{Spectra}, our best-fit results required the addition of a thermal component (either a single-temperature blackbody or multicolor disk blackbody) to a power law Comptonization spectrum with resulting reflection. Curiously, we find an unusually high inner disk temperature of $\gtrsim 3.0$ keV (see Table~2), which is unexpected because the temperature of the thermal component in NS LMXB spectral models (either a blackbody or multicolor disk blackbody, and in many cases both) rarely approaches 3 keV. This is true when modeling reflection features \citep{Cackett_etal2009b, Cackett_etal2010, Sleator_etal2016, Ludlam_etal2017a} as well as when reflection is not detected \citep{Church_etal2010, Church_etal2012, Church_etal2014, Homan_etal2018}. However, based on the available spectra, we cannot rule out such high disk temperatures, and therefore we include the disk model throughout our analysis as an alternative to our model with a single-temperature blackbody. Such a comparison also highlights the dependency of the lag-energy spectrum on whether or not thermal reverberation from the accretion disk is taken into account.

Another complication arises in considering the physicality of the single-temperature blackbody model, since reflection requires an accretion disk, and thermal radiation from that disk should be visible in the time-averaged spectrum. This could be resolved if the disk temperature and normalization are sufficiently low enough that the effect of the disk above 3 keV is minimal.

In each QPO frequency group we then calculated the lag-energy spectrum by convolving the reflected flux fraction from spectral fitting with a two-dimensional transfer function. The resulting lag-energy spectrum depends strongly on whether a blackbody or disk represents the thermal continuum in our spectral model, because thermal reverberation from the disk can be a result of reflection, and we therefore include the disk emission in our reflected flux. This has the result of diluting other reflection features that may be present in the lags.

From the results of fitting the time-averaged spectra, we were able to calculate the reflected flux fraction at all energies, which when convolved with the two-dimensional transfer function describes the timing response for 4U~1728$-$34 at all energies in each QPO frequency group. We generate lag-energy spectra for each of the best-fit spectral models for each QPO frequency group, and compare directly to the lag-energy spectra previously measured for the upper kHz QPOs in 4U~1728$-$34 by \citet{Peille_etal2015}. This comparison is shown in Figure~\ref{lag_data}.

Considering the shape of the lag-energy spectra, we find that the results vary greatly depending on whether or not thermal reverberation from an accretion disk is included in our model (see Fig.~\ref{lag_model}). Models utilizing a disk as the thermal component and therefore including thermal reverberation produce a broad, almost flat lag-energy curve which are either consistent with hard lags or decrease at higher energies. Hardly any distinct features are seen, and these lag-energy spectra lack the characteristic features of reflection --- a broad Fe line and Compton hump above 10 keV. Models including a single-temperature blackbody, on the other hand, produce lag-energy spectra dominated by those very reflection features, decreasing at low energies until $\sim 9$ keV, before showing a hard lag at higher energies. That `v' shape is punctuated by the broad Fe line signature between 6 and 7~keV. Unfortunately, neither of these shapes well represent the lag-energy data previously measured for 4U~1728$-$34, which have a roughly flat spectrum until $\sim 10$ keV, before increasing to show a hard lag. In the case of the 1000--1100 Hz QPO frequency group, a slight bump is seen between 5 and 7 keV, which could represent an Fe line and may therefore suggest reverberation as the cause. In our blackbody models, however, the 1000--1100 Hz bin shows the weakest Fe line.

To further test whether the measured lag-energy spectra might be fit by our reverberation models, we allowed for the possibility of scaling the lags by a constant factor in order to better match the data. In doing so we test whether our methods of introducing dilution might be responsible for the discrepancy between our modeled lags and the data. For a scaling factor greater than one, the prominent Fe line from the blackbody model became problematic even as the increase in the lags at higher energies were better fit, and the $\chi^2$ difference between the models and the data were not improved. In the case of the disk model, a key assumption was that the full flux from the disk responds to variations in the incident flux as thermal reverberation, however this may not be the case. We therefore produced a lag-energy spectrum for each QPO frequency group with $1\%$, $10\%$, or $50\%$ of the disk's flux included in the reflected flux fraction. When a smaller fraction of the disk's flux is included as reverberation, the key reflection features (such as the Fe line and Compton hump) become more pronounced, and the lag-energy spectrum resembles the results of the blackbody model. While in some cases this provided a small improvement in the fit to the measured lags of 4U~1728$-$34, these fits were still poor due to the shape of the modeled lags. Since it is the shape of our lag models that do not match the data, we conclude that reflection and the resulting reverberation cannot fully explain the lag-energy spectra of the upper kHz QPOs in 4U~$1728-34$.

Considering the difference in shape between our reverberation models and the measured lag-energy spectra, we considered whether inverting the shape of the models might better fit the measured lags. This is physically possible under the conditions of phase-wrapping, which has the effect of inverting the lags (see Section~\ref{Model}). Phase-wrapping occurs at the upper kHz QPO frequencies for a canonical 1.4 $M_\odot$ neutron star due to reverberation if the source height is sufficiently high, and we test whether increasing the source height to h = 60 $R_g$ is able to reproduce the measured lags. With the disk model, which accounts for thermal reverberation, we find the best results yet, and still our fits are statistically poor (Table~3). In the case of the blackbody model, our fits are worse than was the case for a lower source height.

It is clear from Figure~\ref{lag_data} that our lag-energy models do not represent the data qualitatively, and this is confirmed by a statistical check --- in the best instances our models fit the data with a reduced $\chi^2~\approx~3$, as shown in Table~3. It is therefore unlikely that reverberation alone causes the lags measured using the upper kHz QPOs in NS LMXBs, which has already been shown to be true for the lower kHz QPOs \citep{Cackett2016}. The recently developed model \texttt{reltrans} \citep{Mastroserio_etal2018, Mastroserio_etal2019, Ingram_etal2019} is a more physically self-consistent reverberation framework, however the lag-energy spectra produced generally matches the overall shape of our lags and also provides a poor fit to the measured lags of 4U~1728$-$34. This suggests that reverberation, regardless of the model, cannot reproduce the measured lag-energy spectra for the upper kHz QPO lags alone.

In the framework discussed above we assume that the fraction of variable flux in each model component has the same dependence on energy, however this is not necessarily true. We tested whether any variable spectrum based on our best-fit spectral models might reproduce the shape of the lags (see Section~4.2). Our best-fit variable spectra still did not match shape of the reflected flux fraction necessary to reproduce the lags, which means that even when allowing the normalizations to vary, reverberation is unable to produce the measured lag-energy spectra of 4U~$1728-34$.

One other limitation of our model is that it assumes a lamppost geometry; that is, it uses a point source of incident X-rays at a given height above the accretion disk. The true physical picture is likely more complex, and may involve an extended corona or some other geometry for the source of hard X-rays, and this could influence the resulting lags from reverberation. Models of extended coronae have been developed for AGN \citep{Wilkins_etal2016}, and it could be that an extended corona model produces a different lag-energy spectrum for the kHz QPOs than we consider here.

\section{Conclusion} \label{Conclusions}

We have modeled the lag-energy spectra for the upper kHz QPOs of 4U~$1728-34$ by fitting the time-averaged spectra in which upper kHz QPOs are significantly detected by \citet{Peille_etal2015}, and then using the reflected flux fraction from our spectral models to produce a two-dimensional transfer function. This represents the first attempt at modeling the upper kHz QPO lags in a NS LMXB using reverberation. We find that the shape of the lag-energy spectrum depends greatly on the choice of continuum model used to describe the time-averaged spectrum, however none of our models accurately represent the lags detected in 4U$1728-34$. It is therefore unlikely that reverberation alone produces the measured lags for the upper kHz QPOs in NS systems.

Since the emission mechanism of both the upper and lower kHz QPOs is not well known, it is possible that the detected lags are produced intrinsically at the source rather than via reverberation with the accretion disk. Some recent models attempting to explain kHz QPOs consider that very possibility, in particular if the QPO variability arises within the Comptonizing boundary region or corona. As thermal photons from the accretion disk or neutron star are up scattered to higher energies, the final energy of a given photon can depend on the number of scatterings, allowing for an energy-dependent time lag \citep[see, e.g.][]{Kumar&Misra2014, Kumar&Misra2016}. The Comptonization model of \citet{Kumar&Misra2014} was recently used by \citet{Ribeiro_etal2019} to model the energy-dependence of both the fractional rms amplitude and the lags of the lower kHz QPOs of 4U~1636$-$53, with particular success in the case of the lags. Even more recently, \citet{Karpouzas_etal2019} reproduced the model by \citet{Kumar&Misra2014} in order to provide encouraging fits to the spectral-timing properties of the kHz QPOs in that same source. Models describing Comptonization remain promising as long as they are able to explain the observed time lags as well as the other spectral-timing properties of the kHz QPOs.

Although modeling the emission and spectral-timing properties of kHz QPOs is complex, and available data on kHz QPOs limited, it remains an important consideration in bettering our understanding of NS LMXBs. It remains a promising topic, because such rapid variability likely depends on the physics of the inner accretion region around the neutron star, so that by understanding QPOs we may gain a powerful tool to measure the physics of some of the most extreme environments available.

\acknowledgements

We thank Adam Ingram and Guglielmo Mastroserio for kindly providing the most recent \texttt{reltrans} model and for discussions on implementing it. BMC and EMC gratefully acknowledge support through NSF CAREER award number AST-1351222.

\software{XSPEC \citep[v12.9.1;][]{Xspec}, reltrans \citep{Mastroserio_etal2018, Mastroserio_etal2019, Ingram_etal2019}}

\bibliographystyle{apj}
\bibliography{apj-jour,benbibliography}

\end{document}